\documentclass[acmsmall]{acmart}

\usepackage{booktabs}
\usepackage{multirow}
\usepackage{subcaption}
\usepackage{xcolor,colortbl}
\usepackage{color, soul}
\usepackage{textcomp}
\usepackage{fontawesome}
\usepackage{longtable}
\usepackage[T1]{fontenc}   

\definecolor{midp}{HTML}{5E96AE}
\definecolor{lowp}{HTML}{BC85A3}

\AtBeginDocument{%
  \providecommand\BibTeX{{%
    \normalfont B\kern-0.5em{\scshape i\kern-0.25em b}\kern-0.8em\TeX}}}

\setcopyright{acmcopyright}
\copyrightyear{2020}
\acmYear{2020}
\acmDOI{10.1145/1122445.1122456}

\acmJournal{POMACS}
\acmVolume{37}
\acmNumber{4}
\acmArticle{111}
\acmMonth{8}

\begin{document}

\title{What Makes People Join Conspiracy Communities?: Role of Social Factors in Conspiracy Engagement}

\author{Shruti Phadke}
\affiliation{%
  \institution{University of Washington}
  \city{Seattle}
  \country{USA}}
\email{phadke@uw.edu}

\author{Mattia Samory}
\affiliation{%
  \institution{GESIS}
  \city{Cologne}
  \country{Germany}}
\email{mattia.samory@gesis.org}

\author{Tanushree Mitra}
\affiliation{%
  \institution{University of Washington}
  \city{Seattle}
  \country{USA}}
\email{tmitra@uw.edu}

\renewcommand{\shortauthors}{Shruti Phadke, Mattia Samory, Tanushree Mitra}

\begin{abstract}
  Widespread conspiracy theories, like those motivating anti-vaccination attitudes or climate change denial, propel collective action and bear society-wide consequences. Yet, empirical research has largely studied conspiracy theory adoption as an individual pursuit, rather than as a socially mediated process. What makes users join communities endorsing and spreading conspiracy theories? We leverage longitudinal data from 56 conspiracy communities on Reddit to compare individual and social factors determining which users join the communities. Using a quasi-experimental approach, we first identify 30K \textit{future conspiracists}---(FC) and 30K matched \textit{non-conspiracists}---(NC). We then provide empirical evidence of importance of social factors across six dimensions relative to the individual factors by analyzing 6 million Reddit comments and posts. Specifically in social factors, we find that dyadic interactions with members of the conspiracy communities and marginalization outside of the conspiracy communities, are the most important social precursors to conspiracy joining---even outperforming individual factor baselines. Our results offer quantitative backing to understand social processes and echo chamber effects in conspiratorial engagement, with important implications for democratic institutions and online communities.
\end{abstract}



\maketitle

\section{Introduction}
The spread of conspiratorial belief and misinformation online is a growing concern. Conspiracy theories of ethnic replacement motivated the mass shootings in El Paso \cite{ElPasosh35online}, Christchurch \cite{Whitegen5online}, and recently Hanau \cite{Hanauatt46online}, which the perpetrators discussed in fringe online communities like 8chan and Gab. Conspiratorial thinking fosters speculations in online discussions and may lead to increased offline consequences, such as the QAnon conspiracy theory about the recent COVID-19 pandemic that drove a train engineer to crash a train near a hospital ship \cite{Conspira76online,Conspira96online}. By its very nature, social media offers a social component to conspiracy discussions where users can interact with each other through discussion threads.  Online conspiracy communities thus bring together multiple heterogeneous groups of individuals with different background beliefs and motivations, sharing similar epistemological concerns  \cite{Klein2018TopicForum}. Once joined, conspiracy community users may radicalize, increasingly engaging with conspiracy and neglecting other communities \cite{Samory2018ConspiraciesEvents}. It is thus crucial to understand the precursors to joining conspiracy communities.

What drives users to join online conspiracy communities? Users who do so, show early on a distinctive use of language and choice of special-interest communities 
\cite{Klein2019PathwaysForum}. This is in line with ample research in social psychology on the \textbf{individual factors} associated with conspiratorial belief \cite{butler1995psychological,goertzel1994belief,hofstadter2012paranoid}. Yet, these studies investigate individuals' attitudes isolated from their social environment. Despite the social nature of conspiracy theorizing online \cite{Franks2013,Stempel2007MediaTheories} and of the collective action it projects onto the real world \cite{Kreko2015CONSPIRACYCOGNITION,mitra2016understanding}, we have surprisingly little insight about the role of \textbf{social factors} in joining online conspiracy communities. This paper provides just such an insight.

We take a socio-constructionist approach---a line of scholarship beholding that meanings are developed in coordination with others rather than separately within each individual \cite{leeds2009social}---and consider online conspiracy discussions as a shared pursuit by a collection of individuals towards making sense of the reality around them. As such, we investigate conspiracy theory adoption as a social phenomena. We leverage the theoretical framework laid out by Sunstein \cite{sunstein2009conspiracy} to identify social factors that may influence conspiracy joining on Reddit---a network of online communities (or subreddits) with dedicated subreddits for conspiracy discussions.  

First, we identify a group of 56 subreddits as \textbf{conspiracy communities} by empirically developing a ``conspiracy scale'' that weighs subreddits from most conspiratorial to most scientific. For example, {\small \tt r/C\_S\_T}, a subreddit that is essentially a sequel to {\small \tt r/conspiracy}---the biggest breeding ground of conspiracies on Reddit---is also most similar to {\small \tt r/conspiracy} on our conspiracy scale. Next, using a retrospective case control study design, we analyze \textbf{future conspiracist (FC)}---Redditors who would go on to \textbf{contribute}---comment or post---in any of the 56 conspiracy communities on Reddit. We implement an intricate statistical matching process to contrast the cohort of \textbf{future conspiracist (FC)} users with a control group of \textbf{non-conspiracist (NC)} users, who never contribute to conspiracy communities but have similar Reddit activity as FC users prior to their \textbf{joining}. Specifically, we compare the direct interactions happening on Reddit threads by FC and NC with \textbf{current conspiracist (CC)}---users currently engaged in conspiracy communities. Based on the direct interactions, we build social factors, such as the preeminence of CC in the users' social circles or social segregation from other communities. Precisely, using Sunstein's framework \cite{sunstein2009conspiracy} we map social factors across six dimensions---availability of conspiracists, informational pressure, reputational pressure, emotional snowballing, group polarization, and self-selection. To provide a reference for the significance of social factors, we also calculate the individual factors related to psychological predisposition \cite{butler1995psychological,goertzel1994belief}, such as feelings of anger, sadness, anxiety and inclinations towards crippled epistemology, \cite{sunstein2009conspiracy}---limited exposure to relevant information.

In all, we compare 30K FC with 30K NC over 6M Reddit contributions using individual and social factors as features in logistic regression.  By analyzing model coefficients in logistic regression, we find ample evidence suggesting that social factors are important towards conspiratorial joining. At least one feature from each dimension is significant. In fact, some social factors have higher predictive power than any of the  individual factors. 
Additional  investigation of the relative importance of different social factors reveal that availability of conspiracists as the most important social factor. In other words, direct exposure to conspiracists and their conspiratorial ideals via direct interactions happening on online platforms is the most important social precursor of conspiracy joining. 

Our results provide us with a unique standing to consider conspiracies as a social effort. This allows us to observe the processes through which conspiracists may experience informational and emotional segregation, face social stigma, and become subject to recruiting efforts by current members of the conspiracy communities. Our results provide evidence of how group polarization and social self selection lead to joining the conspiracy communities. Specifically, we make the following contributions:
\begin{itemize}
    \item Using a data driven approach, we construct the ``conspiracy scale''  to identify conspiratorial subreddits (Figure \ref{fig:ccflow} (2)). The conspiracy scale allows us to characterize subreddits according to their similarity to {\small \tt r/conspiracy} and diversity of user contributions across different subreddits (Section \ref{nonconsp}). 
    \item To study social factors as precursors to conspiracy joining, we undertake an elaborate statistical matching scheme that finds similar future cospiracists and non-conspiracists based on their Reddit joining time, contribution volumes across different time spans, and semantic similarity between the subreddits they contribute in (Figure \ref{fig:rthread}). 
    \item We offer a systematic operationalization of theoretically-motivated individual and social factors towards conspiracy engagement (Section \ref{conspfactors}, Table \ref{tab:summfactor}). 
    \item Through a quasi-experimental study, we detail the individual and social factors correlated with users joining of the conspiracy communities (Section \ref{socimport}). We further assess the relative importance of different social feature groups (Section \ref{reldocimport}) and test the generalizability of social features towards topic-specific and general conspiracy discussion joining (Section \ref{robust}). 
\end{itemize}

To our knowledge this is the first study to establish empirical evidence supporting the importance of social factors in conspiratorial engagement. Overall, our study has implications in content moderation suggesting that excessive and mindless censorship can drive people towards conspiracy communities. Moreover, our results support a socio-constructionist view of conspiracy theorizing and open up future research avenues for studying information mobilization and collective action resulting from conspiracy discourse online. 

\begin{figure*}[t]
    \centering
    \includegraphics[width=0.89\textwidth]{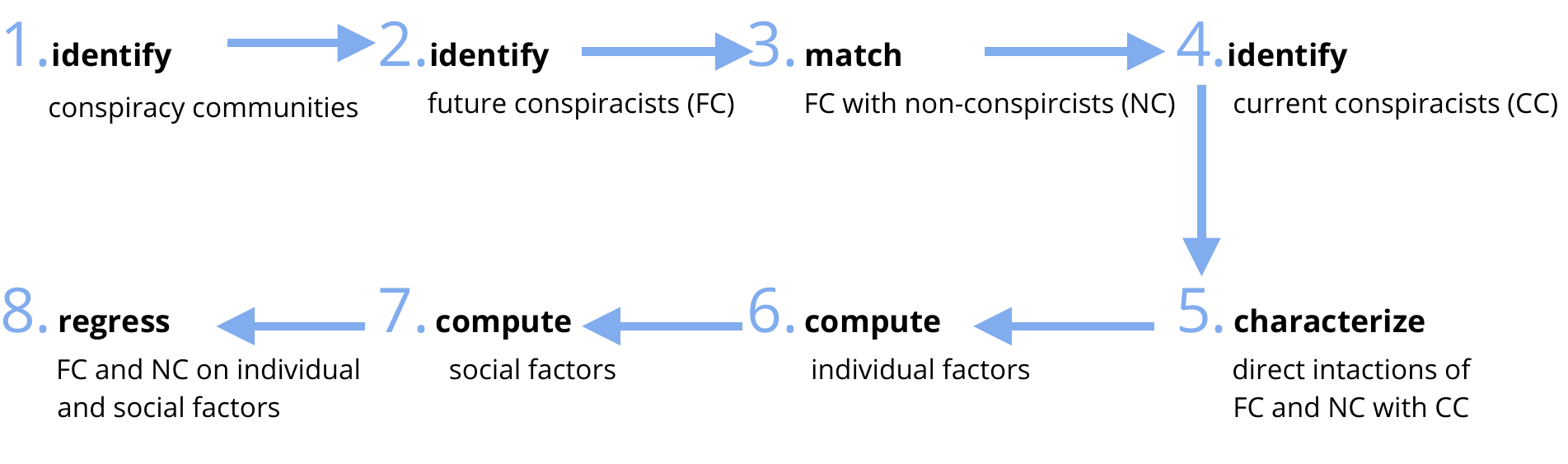}
    \caption{Flowchart detailing our quasi-experimental design and analysis for investigating social factors in conspiratorial joining. We first identify the conspiracy communities (step 1) and then create the cohort of future conspiracists (FC) who will go on to contribute in the conspiracy communities (step 2). We use statistical matching to find non-conspiracists (NC)---users that never contribute in the conspiracy communities but have similar Reddit activity as FC (step 3). In step 4, we identify current conspiracists (CC)---users who are currently engaged in conspiracy communities and characterize the dyadic interactions of FC and NC with CC (step 5). Next, we compute features to capture individual predisposition of FC and NC towards conspiratorial thinking (step 6) and also compute social factors based on dyadic interactions of FC and NC with CC (step 7). Finally, we perform regression analysis using both, individual and social factors and assess the importance of social factors in FC's conspiracy joining (step 8). }
    \label{fig:highmethod}
\end{figure*}

Figure \ref{fig:highmethod} outlines the high-level experimental design and analyses undertaken in this work. The rest of the paper is organized as follows. First, we describe relevant scholarly work on conspiratorial belief. Then we explain our experiment setup and the process of user cohort selection. Next, we describe our operationalization of individual and social factors, followed by a regression analysis. Finally, we discuss the relevance of social factors in social processes in the conspiracy communities, before concluding with the implications and limitations of our findings.

\section{Related work}
Be it vaccine skeptics or in climate change denialists, belief in conspiracy theory fuels collective action and has widespread consequences for society as a whole. Yet, studies of conspiracy theory adoption focused on precursors that are intrinsic to an individual, rather than influenced by the individual's social context. Drawing from literature in collective action and social constructionism we take a step in latter direction and study how social factors influence online users' participation in conspiracy theory communities. Next, we outline existing research on individual factors in conspiracy theory adoption. Then, we describe social aspects of conspiracy adoption. Finally, we review existing work exploring online conspiracy theories.
.

\subsection{Individual Factors in Conspiratorial Belief}
Conspiracy theories are attempts to explain the occurrence of an event as a covert plot orchestrated by secret organizations \cite{banas2013inducing}. 
Research on conspiracy theory adoption have largely focused on individual's psychological and epistemological characteristics \cite{sunstein2009conspiracy}. For example, feelings of hopelessness, insecurity, anxiety, and lack of trust are considered important towards forming conspiratorial beliefs \cite{butler1995psychological,goertzel1994belief}.
Moreover, individuals that engage in conspiratorial beliefs are reported to show characteristics of paranoia \cite{hofstadter2012paranoid}, suspicion towards authoritative information sources \cite{sunstein2009conspiracy} and tendency to believe unsubstantiated or false claims \cite{mitra2016understanding}. 
Previous studies stress that the need for justifying or explaining events forms the very foundation for conspiratorial thinking \cite{mccauley1979popularity,keeley1999conspiracy,vitriol2018illusion,van2018connecting}. For example, \citeauthor{van2018connecting} found that people feel the need to detect patterns or ``connect the dots'' in order to make sense of the physical and social environment they live in \cite{van2018connecting}. This may explain the core process in developing irrational beliefs where people attempt to detect patterns for random events. \citeauthor{sunstein2009conspiracy} suggest that it is thus important to understand how people acquire information related to conspiracies \cite{sunstein2009conspiracy}. Specifically, absence of relevant and ample information can result in ``crippled epistemologies.'' In other words, people who are exposed to very limited relevant information and if what they know is wrong, they have a high likelihood of fixating on their inaccurate beliefs \cite{sunstein2009conspiracy}. 
Though the individuals who believe in conspiracies might be psychologically predisposed, 
their social environment can also play a key role in conspiracy adoption. Thus, in this work we extract cues related to an individual's psychological attributes (e.g., feelings of anger, anxiety, sadness \cite{butler1995psychological,goertzel1994belief}) and their epistemological inclinations (crippled epistemologies \cite{sunstein2009conspiracy}) from their social media activity. These serve as a strong baseline to understand the ability of social factors in identifying which users will join conspiracy theory communities.

\subsection{Social Aspects of Conspiracy Theory Adoption}
In social sense, the development of conspiracy theories can be described by groups of individuals jointly constructing the understandings of the world on the basis of shared identity \cite{Franks2013}. 
From this socio-constructionist stance, conspiracy theories are born from the social processes of filtering available information and deliberating on whether it is true. For example, conspiracy theories prosper in the wake of dramatic events \cite{Samory2018ConspiraciesEvents,Starbird2017} when available information is insufficient to assess its truthfulness. Thus in such situations conspiracy theorizing is an attempt of collective sensemaking \cite{Kou2017ConspiracyMedia}. Studies focusing on the collective processes of consuming information in the context of fake news present crucial insight on the collective pitfalls that may lead to formulating (false) conspiracy theories \cite{Leavitt2017TheGatekeeping,Kou2017ConspiracyMedia}. 
In this work, we explicitly abstain from assessing the truth of conspiracy theories. Our focus, instead, is on the social factors that lead users to conspiracy theory discussions in the first place.

One challenge in studying such social factors in conspiracy theory adoption is the lack of longitudinal data and of granular information of social interactions, prior to individuals' adoption of conspiracy theories. We overcome this challenge by comparing online traces of 30k future conspiracists (FC) before they join conspiracy discussion communities with 30k non-conspiracists (NC) who never join conspiratorial communities on Reddit.  

\subsection{Conspiracies in Social Media}
Research in analyzing conspiracies online mainly explores the effects of exposure to conspiratorial discussions and their linguistic attributes. Specifically,  how conspiratorial belief affects user's information retrieval habits \cite{koutra2015events,coady2006conspiracy}, what causes conspiratorial predisposition \cite{mitra2016understanding,uscinski2016drives}, how individuals discuss conspiracy theories in social media \cite{Starbird2017}, and what linguistic mechanisms are at play in conspiratorial narratives \cite{Samorycscw2018}. However, there is limited empirical evidence as to what leads to the formation of conspiracy theory groups. The present work fills this gap. Perhaps closest to ours, recent work by Klein \cite{Klein2019PathwaysForum} investigates language use and posting patterns of users before they join one conspiracy subreddit on Reddit. In this work, we focus on social, rather than linguistic factors that influence users joining multiple conspiracy communities. Specifically, we adhere to Sunstein's categorization of social features in conspiracy theory adoption \cite{sunstein2009conspiracy}. We describe Sunstein's framework in Section \ref{conspfactors} while elaborating on our feature construction process. Table \ref{tab:summfactor} presents the overview of the individual and social factors and the features derived from them.

\section{The Conspiracy Communities}
\begin{figure*}[t]
    \centering
    \includegraphics[width=0.99\textwidth]{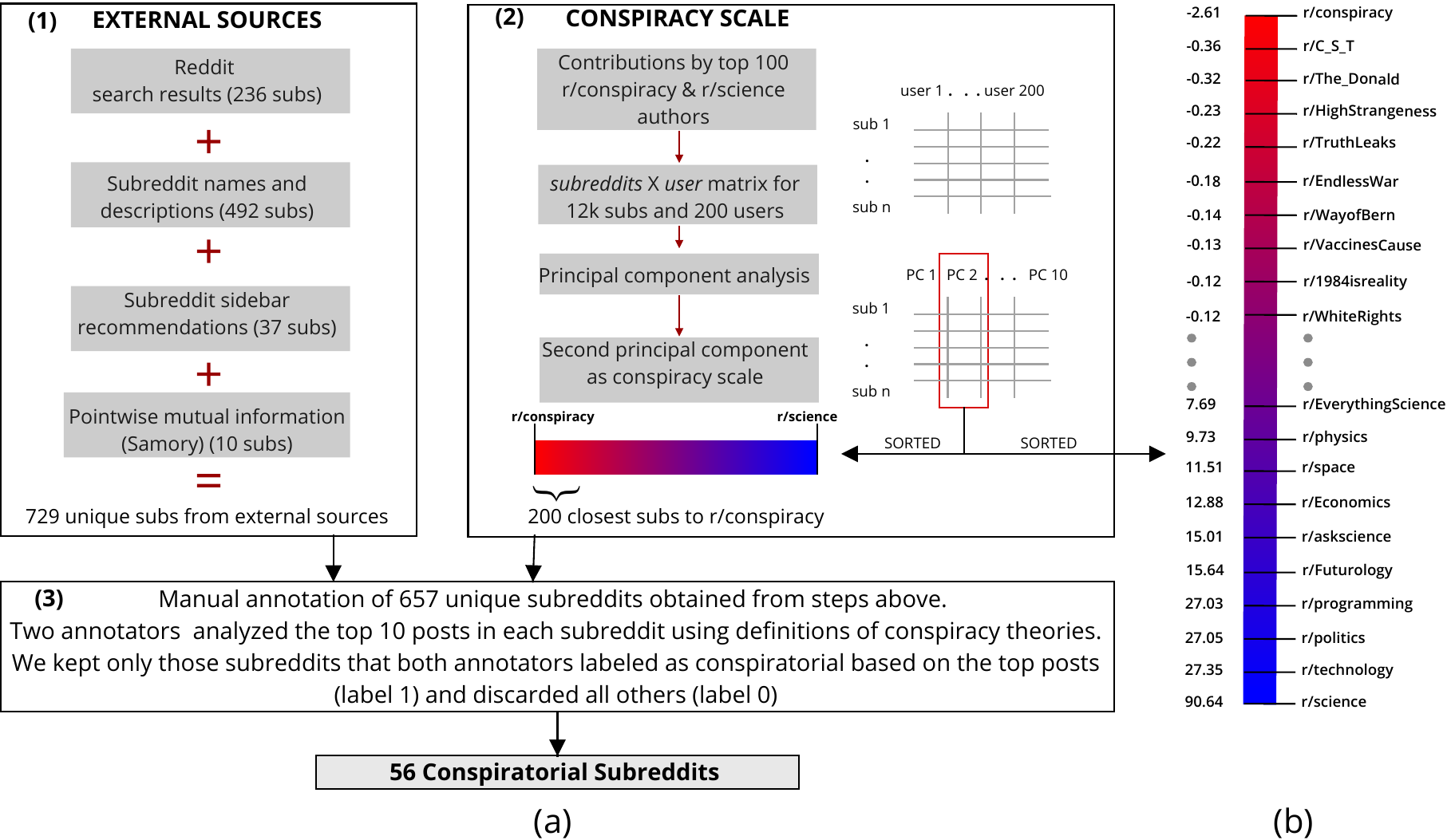}
    \caption{ (a) Flowchart illustrating the process for identifying the conspiracy communities. We obtain the subreddit candidates for conspiracy communities using both, (1) external sources and (2) conspiracy scale. The conspiracy scale is generated by sorted second principal component of subreddit $\times$ user matrix of 
    top {\small \tt r/conspiracy} and {\small \tt r/science} user contributions across different subreddits. We take leftmost 200 (200 subreddits closest to {\small \tt r/conspiracy}) from conspiracy scale along with subreddits from external sources as candidates for conspiracy communities. We manually annotate the candidate subreddits to include 56 subreddits in conspiracy communities   (b) Top 10 subreddits on both sides of the conspiracy scale alongwith their weights according to the 2nd principal component. We did not normalize the weights to preserve sparsity in values of the principal component. }
    \label{fig:ccflow}
\end{figure*}

We take a socio-constructionist stance, and consider the collective of users producing conspiracy discussions as a community producing knowledge. Specifically on Reddit, we first define a group of subreddits hosting conspiratorial discussions as the ``conspiracy communities''. Identifying subreddits that engage in conspiracy discussions is a challenging process for several reasons. Reddit has a total of 1.2 million subreddits with no global taxonomy that could help us easily understand the themes in different subreddits. Previous researchers have commonly focused only on {\small \tt r/conspiracy}--a subreddit dedicated to discussing all types of conspiracy theories---to study conspiratorial engagement and narratives \cite{Samory2018ConspiraciesEvents,Samorycscw2018,Klein2019PathwaysForum}. However,  there are several other communities on Reddit that promote conspiracy theories as well. In that, some subreddits openly self-identify as conspiracy discussion communities  while others host conspiratorial content without having it as their primary focus. For example, {\small \tt r/ConspiracyII} invites only conspiracy theories whereas {\small \tt r/ConspiracyNews} focuses on reporting news around conspiratorial topics. 
Moreover, even within the solely conspiratorial communities, some specialize on just one or few related conspiracy theory narratives and others welcome all types of discussions. To elaborate, there are specialized subreddits dedicated to discussing specific  conspiracies, such as moon landing hoax and flat earth ({\small \tt r/moonhoax} and {\small \tt r/theworldisflat}, respectively) while others welcome any and all kinds of conspiratorial discussions ({\small \tt r/FringeTheory} and {\small \tt r/ConspiracyZone}).  Given such high diversity in conspiracy discussion subreddits, it is imperative that we identify conspiracy communities with high precision. Towards this end, we employ a multi-stage, mixed-methods approach to first, mine potentially conspiratorial subreddits and then, carefully vet them using human judgement. Specifically, to find the candidates for conspiracy communities, we resort to two key steps. First we look at external sources such as Reddit recommendations and methods based on previous research. 
Second, we devise a conspiracy scale that weighs subreddits based on their similarity to {\small \tt r/conspiracy}. Figure \ref{fig:ccflow} displays the entire process of identifying the conspiracy communities.

\subsection{External Sources: }
We look at four external sources for finding conspiratorial communities. Specifically, Reddit search results, subreddit names and descriptions, subreddit sidebar recommendations and mutual information based methods used in previous research. Table \ref{tab:sub_examples} provides examples for subreddits mined from each of the four external sources alongwith the conspiracy scale described later. 
\begin{enumerate}
    \item \textbf{Reddit search results: } We look at search results provided by Reddit to emulate how Reddit users might find conspiratorial subreddits.
    Reddit has a search bar at the top of any page in which users can enter the search query to find different subreddits, users and posts. 
    We search the term `conspiracy' on Reddit's home page and note 236 recommended subreddits. 
    \item \textbf{Subreddit name and description: } We consider the user's choice of knowingly participating in conspiracy discussions as an important criterion towards identifying Redditors that engage in conspiracies. Reddit users can understand the theme of a subreddit by subreddit names or the descriptions. Hence, we refer to the subreddit names and descriptions available on  
    {\small \tt files.pushshift.io/reddit} \footnote{\url{https://files.pushshift.io/reddit/} is a publicly accessible repository of Reddit datasets maintained by  Jason Baumgartner \cite{baumgartner2020pushshift}}. Since we are interested in selecting self-identifying conspiracy subreddits, we perform regular expression match for the string ``conspir'' in the descriptions and the names. 
    \item \textbf{Subreddit sidebar recommendations: } Often, subreddit descriptions contain a sidebar in which the other related subreddits are listed. For example, {\small \tt r/conspiracy} lists {\small \tt r/Wikileaks} and {\small \tt r/Endlesswar} as ``related subreddits'' it the sidebar (Table \ref{tab:sub_examples}).  Hence we looked at sidebar recommendations for subreddits obtained in step 2. We continued this process recursively until there were no more sidebar recommendations or the recommended subreddits were already listed in step 1-3. This process resulted in 37 new subreddits. 
    \item \textbf{Pointwise mutual information: } We also look at the work by other researchers characterizing conspiratorial communities. Specifically, Samory et. al. \cite{Samory2018ConspiraciesEvents} find communities that share surprising number of common users to {\small \tt r/conspiracy} (page 5, Table 1 in \cite{Samory2018ConspiraciesEvents} ). We consider top 10 subreddits listed in \cite{Samory2018ConspiraciesEvents} as potentially conspiratorial subreddits. 
\end{enumerate}

We look at multiple sources for identifying conspiratorial subreddits to select conspiracy communities with high precision. While external sources produce useful candidates for the conspiracy communities, they are also limited in their effectiveness. Reddit recommendations produce high number of irrelevant suggestions. For example, we find several subreddits unrelated to conspiracy even within the top 10 search results on Reddit ({\small \tt r/todayilearned} Table \ref{tab:sub_examples}). In addition, subreddit sidebars are not always populated by the subreddit community. 
Moreover, pointwise mutual information approach extracts subreddits that are distinctively similar to {\small \tt r/conspiracy} favouring smaller subreddits. For example, all of the top 10 subreddits closer to {\small \tt r/conspiracy} listed in \cite{Samory2018ConspiraciesEvents} have less number of subscribers and contribution volume (Table \ref{tab:sub_examples}). Hence, we look for a data-driven, scalable approach that could capture subreddits that are generally, and not just surprisingly, similar to {\small \tt r/conspiracy}. Specifically, we perform Principal Component Analysis (PCA) on contributions received in subreddits by different users to devise the ``conspiracy scale'' (Figure \ref{fig:ccflow} a and b). 

\subsection{Conspiracy scale}
We design the conspiracy scale to characterize semantic similarity between subreddits in terms of shared user base \cite{martin2017community2vec}. 
Previously, other researchers have also employed user participation based measures to compare subreddits \cite{martin2017community2vec,Kumar2018CommunityWeb,Waller2019GeneralistsPlatforms}. However, such representations are not designed to specifically study the conspiratorial nature of the subreddits. Samory et. al. \cite{Samory2018ConspiraciesEvents} use  pointwise mutual information (PMI) to identify communities that are distinctively similar to {\small \tt r/conspiracy}. While their approach successfully identifies conspiratorial subreddits, it focuses on identifying communities that share \emph{surprisingly} common users with {\small \tt r/conspiracy}. To elaborate, PMI is a co-occurrence based measure where mutual information between two subreddits is calculated based upon the number of common users in them. PMI can return \emph{surprisingly} similar subreddits to {\small \tt r/conspiracy} because it is known to be biased towards low frequency or rare items (or subreddits in this case) \cite{bouma2009normalized,kaminskas2014measuring}. To bridge this gap, we search for the method that would not be biased towards just surprise while measuring similarity between the subreddits.   
We take intuition from Samory et. al. \cite{Samory2018ConspiraciesEvents} specifically that conspiratorial subreddits can be identified by contrasting the user activity in {\small \tt r/conspiracy} to its polar opposite community---{\small \tt r/science} \cite{Bessi2015ScienceMisinformation}. However, instead of focusing on finding subreddits that are distinctively similar to {\small \tt r/conspiracy}, we try to understand the similarity based on the variance in user-subreddit participation for users in  {\small \tt r/conspiracy} and {\small \tt r/science} based on Principal Component Analysis (PCA).  

\subsubsection{Creating conspiracy scale} Figure \ref{fig:ccflow} illustrates the process of devising the conspiracy scale. 
Previous scholars have shown that people who believe in one conspiracy tend to believe in others as well \cite{Bessi2015TrendMisinformation,Mitra2016UnderstandingMedia,Starbird2017}. 
Accordingly, we presume that top contributing users---users with highest number of contributions---from {\small \tt r/conspiracy} will have propensity to engage in other conspiracy related subreddits. Further juxtaposing their activity with top contributing {\small \tt r/science} users can help enhance the contrast between conspiratorial and scientific subreddits.  Hence, after removing bot accounts, we select the 100 top contributing users from each of the two subreddits. We extract the entire contribution timelines of these users across all subreddits using {\small \tt pushshift.io} \cite{baumgartner2020pushshift}. In all, this starting dataset spans over 12k subreddits. One could understand the variance in types of the 12k subreddits by analyzing the number of contributions made by the users in each of those subreddits. For example, just by sorting the raw counts of contributions within each subreddit, one could distinguish subreddits with larger subscriber counts from the smaller ones. For the task at hand, we want to extract the directionality in subreddits that places them from most similar to {\small \tt r/science} to most similar to {\small \tt r/conspiracy}. Principal Component Analysis (PCA) is a dimentionality reduction technique that could reduce the data along principal components that explain the maximal amount of variance. Intuitively, the first few components should give us different viewpoints to understand the variance in types of subreddits. Hence, we construct  a $subreddit \times user$ matrix with values indicating contributions made by a user (column of the matrix) in a subreddit (row of the matrix) and apply PCA on it. Specifically, we extract first 10 components ranked based on the amount of variance they explain. Our underlying assumption here is that {\small \tt r/conspiracy} users engage with more conspiratorial subreddits while {\small \tt r/science} users engage with non conspiratorial subreddits. Hence we look for the principal component that projects subreddits in a way that places conspiratorial subreddits on one end and non-conspiratorial subreddits (subreddits similar to {\small \tt r/science}) on the other end, resulting in maximal variance. The first component arranged the subreddits from smallest to largest---summarizing the general variety between subreddits. However, when sorted by \emph{second} component of the PCA (Figure \ref{fig:ccflow}(a)(2)), {\small \tt r/science} and {\small \tt r/conspiracy} fall on two extreme ends indicating that the second component explains the second order variance that the first component does not capture. Since the second component identifies two poles in subreddits---{\small \tt r/science} and {\small \tt r/conspiracy}, we use it as the conspiracy scale. We consider top 200 subreddits from the conspiracy scale as candidates for the conspiracy communities (Figure \ref{fig:ccflow} (b)).

\begin{table}
\centering
\resizebox{\textwidth}{!}{%
\begin{tabular}{@{}llllll@{}}
\toprule
\rowcolor[HTML]{C0C0C0} 
\textbf{} & \textbf{} & \textbf{} & \textbf{Our conspiracy scale} & \textbf{PMI \cite{Samory2018ConspiraciesEvents}} & \textbf{Subreddit embeddings \cite{Kumar2018CommunityWeb}} \\ \midrule
\cellcolor[HTML]{C0C0C0} & method &  & \begin{tabular}[c]{@{}l@{}}conspiracy scale sorted \\ from r/conspiracy to r/science\end{tabular} & \begin{tabular}[c]{@{}l@{}}ranked by closeness \\ to r/conspiracy by PMI\end{tabular} & \begin{tabular}[c]{@{}l@{}}ranked by cosine distance \\ to  r/conspiracy vector\end{tabular} \\
\cellcolor[HTML]{C0C0C0} &  & \textit{rms} & \textbf{3332} & 4271 & 17,741 \\
\multirow{-3}{*}{\cellcolor[HTML]{C0C0C0}\textbf{\begin{tabular}[c]{@{}l@{}}Conspiracy ranks for \\ conspiracy related subreddits\end{tabular}}} & \multirow{-2}{*}{results} & \textit{std} & \textbf{4447} & 4939 & 11,567 \\ \midrule
\cellcolor[HTML]{C0C0C0} & method &  & \begin{tabular}[c]{@{}l@{}}conspiracy scale sorted \\ from r/science to r/conspiracy\end{tabular} & \begin{tabular}[c]{@{}l@{}}ranked by closeness \\ to r/science by PMI\end{tabular} & \begin{tabular}[c]{@{}l@{}}ranked by cosine distance\\  to r/science vector\end{tabular} \\
\cellcolor[HTML]{C0C0C0} &  & \textit{rms} & \textbf{3570} & 4385 & 14,786 \\
\multirow{-3}{*}{\cellcolor[HTML]{C0C0C0}\textbf{\begin{tabular}[c]{@{}l@{}}Science ranks for \\ science related subreddits\end{tabular}}} & \multirow{-2}{*}{results} & \textit{std} & \textbf{5816} & 7102 & 21,448 \\ \bottomrule
\end{tabular}%
}
\caption{ We validate the conspiracy scale generated from PCA (Figure \ref{fig:ccflow} (a)(2) and (b)) by comparing the ranks generated for conspiracy and science related subreddits by our conspiracy scale and other approaches (\cite{Kumar2018CommunityWeb,Samory2018ConspiraciesEvents}). This table  describes the methods used for generating ranks and the results (root mean square and standard deviation) for the ranks obtained. Our method has lowest rms and std in both, conspiracy and science rankings indicating that out scale places conspiracy related subreddits closer to r/conspiracy and science related subreddits closer to r/science. }
\label{tab:scale_valid}
\end{table}

\subsubsection{Validating conspiracy scale}
How well does our scale place conspiracy related subreddit on conspiracy end and science related subreddits on the other? How does it compare with other subreddit similarity measures? We compare our conspiracy scale with two external subreddit similarity measures---pointwise mutual information by Samory et. al. \cite{Samory2018ConspiraciesEvents} and community embeddings by Kumar et. al. \cite{Kumar2018CommunityWeb}.
First, we generate the list of (i) conspiracy related and (ii) science related subreddits based on the subreddit names and descriptions. Specifically, we search for the substring ``conspir'' and ``sci'' in subreddit names and terms ``conspiracy''. ``conspiracies'', ``science'' and ``scientific'' in subreddit descriptions. We curate this list to keep only relevant subreddits in both (i) and (ii). 
Next, we rank the subreddits using the three methods as described in Table \ref{tab:scale_valid}.  In all three methods, {\small \tt r/conspiracy} and {\small \tt r/science} have rank 1 in conspiracy and science ranks respectively. Thus it follows that the conspiracy ranks for conspiracy related subreddits should be close to one. Similarly, the science ranks for science related subreddits should be close to one. Hence, to compare the aggregate ranking of subreddits across all three methods, we calculate root mean square and standard deviation of the ranks generated. In both, conspiracy and science, our scale produces lower standard deviation and root mean square (rms) in the rankings (See Table \ref{tab:scale_valid}) indicating that out scale places conspiracy related subreddits closer to {\small \tt r/conspiracy}  and science related subreddits closer to {\small \tt r/science}. For examples, see top 10 subreddits on both sides of the conspiracy scale \ref{fig:ccflow} (b).  Moreover, unlike the similarity generated by the pointwise mutual information, our scale is not biased towards smaller or larger subreddits. For example, top 10 subreddits closes to {\small \tt r/conspiracy} on the conspiracy scale (Figure \ref{fig:ccflow} (c)) contain both, smaller and larger subreddits with respect to the subscriber count and the contribution volume.

\begin{table*}[t]
\centering \sffamily
\resizebox{\textwidth}{!}{%
\begin{tabular}{@{}llrll|l@{}}
\toprule
\rowcolor[HTML]{C0C0C0} 
\multicolumn{5}{c|}{\cellcolor[HTML]{C0C0C0}\textbf{External Sources}} & \multicolumn{1}{c}{\cellcolor[HTML]{C0C0C0}\textbf{Conspiracy Scale}} \\ \midrule
\textbf{Reddit search results} & \textbf{Subreddit names \& descriptions} & \multicolumn{2}{c}{\textbf{Subreddit Sidebar Recommendations}} & \textbf{PMI \cite{Samory2018ConspiraciesEvents}} & \textbf{} \\
\textbf{r/conspiracy} & r/thoseconspiracyguys &  & r/Wikileaks & r/CHEMPRINTS & r/conspiracy \\
r/insanepeoplefacebook & r/conspiratard &  & r/EndlessWar & r/bilderberg & r/C\_S\_T \\
r/WTF & r/muaconspiracy &  & r/PostCollapse & r/conspiracyhub & r/The\_Donald \\
r/911truth & r/conspiracyundone & \multirow{-4}{*}{r/conspiracy} & r/Documentaries & r/greenlight2 & r/HighStrangeness \\ \cmidrule(lr){3-4}
r/PanicHistory & r/ConspiracyMemes & r/conspiracytheories & r/skeptic & r/WhiteNationalism & r/TruthLeaks \\
r/TrueReddit & r/conspiracies & \multicolumn{1}{l}{} & r/spacex & r/greenlight & r/EndlessWar \\
r/WikiLeaks & r/ConspiracyII & \multicolumn{1}{l}{} & r/fakenews & r/HealthConspiracy & r/WayofBern \\
r/isconspiracyracist & r/actualconspiracies & \multicolumn{1}{l}{} & r/bigfoot & r/OccupyLangley & r/VaccineCause \\ \cmidrule(lr){3-4}
r/todayilearned & r/pokemonconspiracies & r/conspiracyII & r/OccultConspiracy & r/mysterybabylon & r/1984isreality \\
r/politics & r/OccultConspiracy & \multicolumn{1}{l}{} & r/TheTranslucentSociety & r/moonhoax & r/WhiteRights
\end{tabular}%
}
\caption{Table listing the example subreddits obtained by each method. In all, we obtained a total of 657 unique subreddits from every method.  }
\label{tab:sub_examples}
\end{table*}

\subsection{Annotating conspiracy communities}
Table \ref{tab:sub_examples} provides examples of subreddits obtained from every method discussed above. 
With the subreddit list obtained from the external sources and the conspiracy scale, we have 657 candidates for the conspiracy communities. For each candidate, we obtained annotations from two separate annotators who had sufficient experience and context for distinguishing conspiratorial and non-conspiratorial discussions. First, the annotators read the subreddit names and their descriptions. Then, they read at least top 10 submissions from each of the subreddits and analyzed them using the  definitions of conspiracy theories aggregated in \cite{Samorycscw2018}. For example, one of the definitions states: ``\textit{...(conspiracies) involve multiple actors working together in secret to achieve hidden goals that are perceived to be unlawful or malevolent...}'' \cite{Abalakina-Paap1999BeliefsConspiracies,Darwin2011BeliefSchizotypy,vanProoijen2013BeliefMorality,Zonis1994}.
The annotators annotated the subreddit as 1 if either of the definitions applied in at least five posts and 0 otherwise (Figure \ref{fig:ccflow}(a)(3)). We discarded all subreddits that either of the annotators found to be irrelevant or anti-conspiracy or were about trolling conspiracists. For example, {\small \tt r/ChickenApocalypse} contains jokes mentioning the conspiracies about chicken controlling the world and {\small \tt r/Disinfo} is a watchdog subreddit for cataloguing misinformation and debunking conspiracy theories. 
After manual validation, we obtained a list of 56 subreddits that both annotators considered to host conspiracy discussions, ensuring high precision.  We provide a list of subreddits in the conspiracy communities alongwith the links to the example posts containing conspiracies in the supplementary material.

\section{Data and Subjects}

Informed by collective action theories, we hypothesize that the participants in the conspiracy communities---\textbf{Current Conspiracists, CC}---exert some influence on people outside of the communities, via discussions and other interactions. In this light, our research question thus investigates if and how the influence of CC leads users to join the conspiracy communities. We measure CC's influence over users who will, at some point, join any one of the conspiracy communities (\textbf{Future Conspiracists, FC}). We compare and contrast CC's influence on FC against a matched cohort of control users. Using statistical matching, we find users that are comparable to FC in all respects but who never join any of the conspiracy communities---\textbf{Non Conspiracists, NC} (Figure \ref{fig:rthread}). Below, we detail the source of our discussion data and our process for selecting our user cohorts FC, NC, and CC, and how we match FC and NC.

\subsection{Discussions on Reddit}
We study conspiracists on Reddit, a social media platform where users can create, share, and discuss content by participating in specific subdivisions of Reddit (or subreddits). Subreddits contain discussions around specific themes. For example, {\small \tt r/Kanye} is for discussing anything related to Kanye West and {\small \tt r/nintendo} is a subreddit for  Nintendo news and games. Discussions in subreddits start with an opening post called submission, that sets the theme. Users can comment on the submissions and on other users' comments. For the sake of simplicity, we collectively call submissions and comments as ``\textbf{contributions}'', and all contributions in a discussion as ``thread''  (see Figure \ref{fig:rthread} (c)).

\subsection{Finding Future Conspiracists (FC)}
\begin{figure*}[t]
    \centering
    \includegraphics[width=0.99\textwidth]{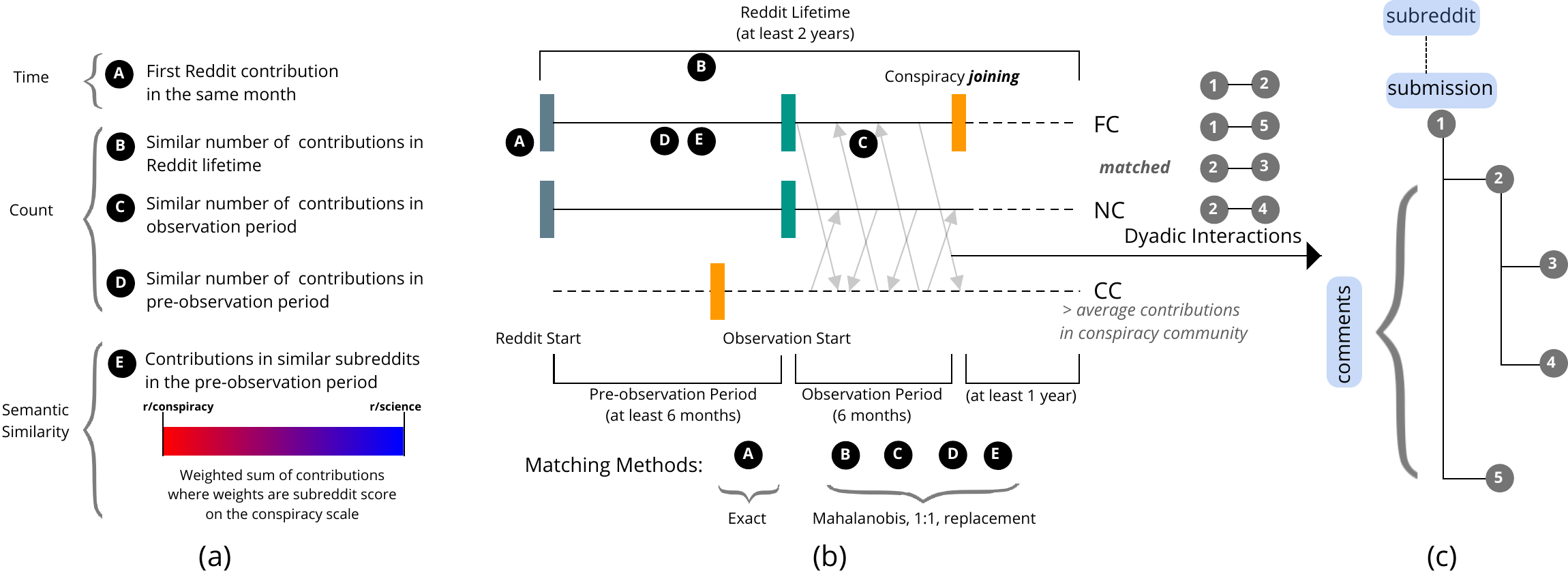}
    \caption{ (a) Matching criteria used for finding similar FC and NC (b) Different time spans in the users' Reddit life. FC and NC are matched on \protect\resizebox{0.35cm}{!}{\protect\includegraphics{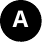}} exactly and on  
    \protect\resizebox{0.35cm}{!}{\protect\includegraphics{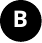}}, 
    \protect\resizebox{0.35cm}{!}{\protect\includegraphics{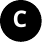}}, 
    \protect\resizebox{0.35cm}{!}{\protect\includegraphics{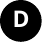}}, 
    \protect\resizebox{0.35cm}{!}{\protect\includegraphics{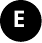}} criteria using nearest neighbor matching with Mahalanobis distance. 
   \protect\resizebox{0.15cm}{12pt}{\protect\includegraphics{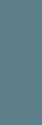}} symbolizes Reddit start,
    \protect\resizebox{0.15cm}{12pt}{\protect\includegraphics{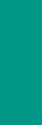}} observation period start and 
    \protect\resizebox{0.15cm}{12pt}{\protect\includegraphics{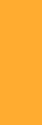}} signifies FC's and CC's joining any one of the conspiracy communities. FC and NC have at least 2 years of Reddit life, at least 10 contributions in the observation and 10 in pre-observation period. We calculate the users dyadic interactions with CC in the observation period. (c) Elaboration of dyadic interactions (faded gray arrows in (a)) indicating replies to or from CC) with examples of dyadic interactions in a typical Reddit discussion thread.  }
    \label{fig:rthread}
\end{figure*}

Figure \ref{fig:rthread} (b) outlines different time spans that we use throughout this paper to characterize users' lifetime on Reddit. FC are Reddit users who eventually engage with any one of the conspiracy communities. We consider the time of their first contribution to any of the conspiracy subreddits as the time when a FC \textbf{joins} the community. We consider the 6 months preceding their joining as the \textbf{observation period} in which we study the individual and social factors affecting their joining. A total of 740,093 users ever contributed to the conspiracy communities; however, we impose a number of constraints to obtain a high quality sample of FC. We want to study FC who become actively engaged, and not users who post only incidentally such as spammers and trolls. Therefore, for each subreddit in the conspiracy communities, we calculate median number of contributions made by users in that community in their lifetime. We consider users that contribute more than the previously calculated median in any of the conspiracy communities as treatment candidates. 
To eliminate throwaway accounts we also remove users with less than 2 years of Reddit lifetime. To reliably measure signals of social factors during the observation period, we keep only users who have enough data---5 contributions---in that time. Finally, in order to reliably match FC and NC, we limit to users with at least 5 contributions and 6 months of activity prior to the observation period. Our final set of FC consists of 30,325 users.

\subsection{Finding Non Conspiracists (NC)}
\label{nonconsp}
To understand the prominence of individual and social factors towards conspiratorial engagement in FC, we need to compare such factors in normal Reddit users, the control group of NC users. Ideally, we want the FC and NC to be indistinguishable based on their Reddit contributions and tenure, but for the fact that NC never join any of the conspiracy communities. We begin with a list of 10 million Reddit users who have at least 2 years of Reddit lifetime and no contributions in the conspiracy communities. Next, we refine this list to match the group of FC users based on the following criteria. 

\begin{itemize}
    \item \textbf{A: } Reddit start month. To select users with similar Reddit tenure, we first match FC with all NC candidates that made first contribution in Reddit in the same month as FC.
\end{itemize}

Next, we want users that are similarly active on Reddit. We define different time spans over FC's life and find NC that have similar contributions in those time periods. Specifically, we match on:

\begin{itemize}
    \item \textbf{B: }Contribution volume in the Reddit lifetime
    \item \textbf{C: }Contribution volume in the observation period
    \item \textbf{D: }Contribution volume in the pre-observation period
\end{itemize}

Finally, we also consider the similarity in contributions during the pre-observation period.

\begin{figure*}[t]
    \centering
        \includegraphics[width=\textwidth]{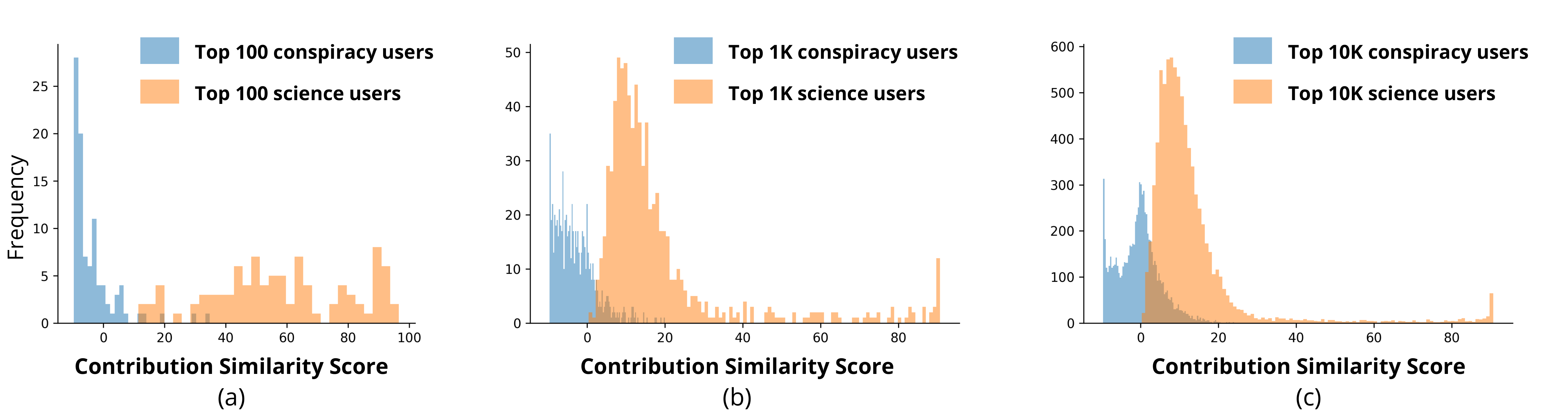}
    \caption{ We obtained the contribution similarity scores (matching criteria \protect\resizebox{0.35cm}{!}{\protect\includegraphics{figures/E.pdf}} ) for (a) top 100 (b) top 1000 and (c) top 10k {\small \tt r/conspiracy} and {\small \tt r/science} users. The scores were generated by taking weighted contributions by users (weights are the conspiracy scale weights for the subreddits). The Wilcoxon rank sum test between distributions returned p-values $<0.05$ in all three cases. This indicates that our contribution similarity calculation is able to characterize different types of users accurately based on their Reddit activity.}
    \label{fig:consci}
\end{figure*}

\begin{itemize}
    \item \textbf{E: } Contributions in similar subreddits in the pre-observation period. We want to control for the types of subreddits FC and NC contribute in, prior to the observation period. Controlling for contributions in similar subreddits can give us FC and NC users who have tendencies to contribute in similar subreddits. We assess the similarity of subreddit contributions made by FC and NC using the conspiracy scale described in Section 3.2. The conspiracy scale gives us weights for subreddits based on their similarity to {\small \tt r/conspiracy} (see Figure \ref{fig:ccflow} b for example). Hence, understanding the users' subreddit activity using conspiracy scale can help us match FC and NC that are similarly conspiratorial or non-conspiratorial in the pre-observation period. Moreover, having FC and NC who have a history of contributions in similar subreddits can give us user cohorts with comparable chances of social interactions with other conspiracists. Thus, to compute our final matching criteria (E), we take weighted sum of normalized user contributions using the subreddit's score on the conspiracy scale. For example, if a user has 60\% contributions in {\small \tt r/C\_S\_T} (-0.36 on scale) and 40\% contributions in {\small \tt r/The\_Donald} (-0.32) then the matching feature value is calculated as $-0.36 \times 0.6 + (-0.32) \times 0.4 = -0.344$. To validate that this feature is able to characterize the users' subreddit contributions effectively, we plot the feature values for top 100, 1000 and 10k  {\small \tt r/conspiracy} and {\small \tt r/science} users and compare their distributions. (See Figure \ref{fig:consci}). In all three cases, Wilcoxon signed rank sum test revealed that the distributions for conspiracy and science users are significantly different which means our contribution similarity calculation is able to characterize different types of users accurately based on their Reddit activity. 

\end{itemize}

\subsection{Matching FC and NC}
For each of the 30K FC, we select one NC from a pool of 3 million NC candidates using statistical matching. Since we want to find FC and NC that join Reddit in the exact same month, we perform exact matching on the Reddit start month criteria. We perform nearest neighbor matching with replacement using Mahalanobis distance on the remaining constraints. 
The matching procedure results in a set of 30,325 FC users matched with  29,098 NC users \footnote{29,098 NC users are mapped to 30,325 FC users because we allow replacements in the matching: to ensure the integrity of the results, we consider different observation periods for all FC and NC user pairs, in effect considering one NC user mapped to two FC users as two distinct NC users}. Note that our matching procedure is more involved than previous empirical studies in conspiracy precursors \cite{Klein2019PathwaysForum} to ensure highly comparable groups of FC and NC users. We use five criteria (Figure \ref{fig:match} (a)) that find similar FC and NC users based on their Reddit joining, volume of contribution across different time periods and also the semantic similarity of subreddits they contribute in. Our intricate matching process that compares different attributes of the users' Reddit activity, enables us to confidently examine the social factors as precursors to conspiracy joining.  

\subsubsection{Quality of matching}
\begin{figure*}
    \centering
    \includegraphics[width=0.52\textwidth]{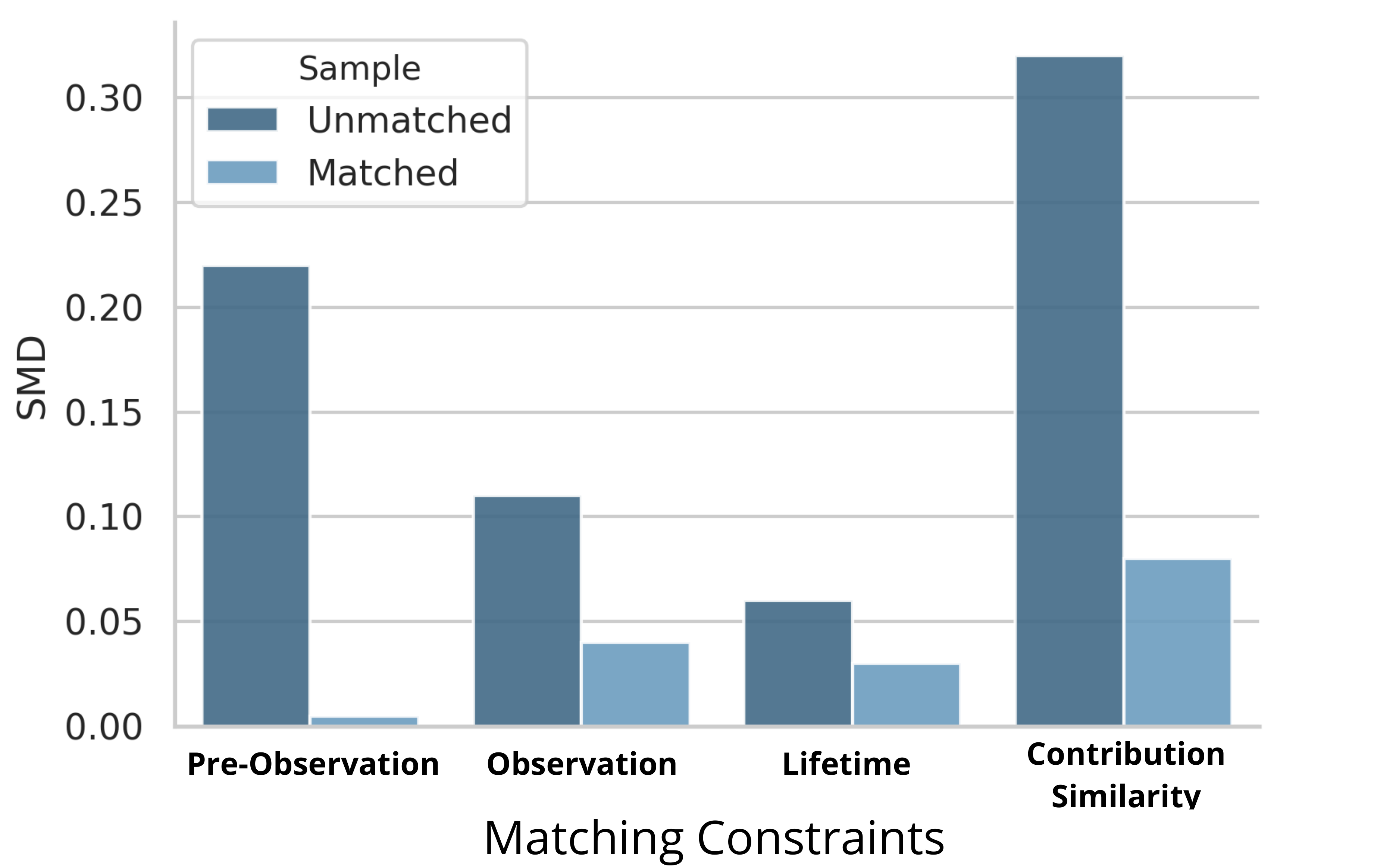}
    \caption{Standardized Mean Difference (SMD) for unmatched and matched users across   \protect\resizebox{0.35cm}{!}{\protect\includegraphics{figures/B.pdf}}, 
    \protect\resizebox{0.35cm}{!}{\protect\includegraphics{figures/C.pdf}}, 
    \protect\resizebox{0.35cm}{!}{\protect\includegraphics{figures/D.pdf}}, 
    \protect\resizebox{0.35cm}{!}{\protect\includegraphics{figures/E.pdf}} matching constraints. All four constraints result in the SMD $<$ 0.08 after matching indicating balanced matched groups of future conspiracists and non conspiracists}
    \label{fig:match}
\end{figure*}

To ensure that our matched FC and NC are statistically comparable, we check the improvement in balance across all of the matching constraints using Standardized Mean Difference (SMD)---a method commonly used by other researchers studying users on social media \cite{saha2019social,sahacausal}. Note that the FC and NC are matched exactly on the first criteria---\resizebox{0.35cm}{!}{\includegraphics{figures/A.pdf}} Reddit start month. Hence, we claculate the SMD for only the rest of the matching criteria. SMD calculates the difference in the means of distributions between the two groups as a fraction of the pooled standard deviation of the two groups. Balanced groups are considered to have SMD less than 0.2 \cite{kiciman2018using}. We obtain an SMD of less than 0.08 across all of the matching constraints, suggesting high quality of matching (See Figure \ref{fig:match}). Specifically, we find 77\% balance increment in pre-observation contributions, 63\% in observation period and 50\% in Reddit lifetime contributions.  We find highest balance improvement in contribution similarity scores (80\%).   

 \subsection{Finding Current Conspiracists (CC)} After matching FC and NC, we have a unique observation window for each matched FC and NC user pair (Figure \ref{fig:rthread}(a)). Studying interactions with conspiracists in the observation period can inform about the social influence the conspiracists have on the conspiracy joining of FC. Hence,  we select a group of users---\textit{current conspiracists (CC)}---that have already joined the conspiracy communities. Each FC and NC has their own set of CC who have already made their first contribution in any of the \emph{conspiracy communities} and who make above average contributions in aggregate in \emph{conspiracy communities} in their lifetime. In other words, for every FC and NC pair, we select their own set of CC based on their unique observation window. 
 In total, there are 61,073 CC involved in the interactions with FC and NC. 
 
 \subsection{Characterizing social interactions with CC} For every FC/NC, we characterize their interactions with CC in the observation period.  Specifically, we look at publicly available interactions between users in Reddit discussion threads. Figure \ref{fig:rthread}(c) demonstrates a discussion thread. We consider ``dyadic interaction'' as a communication between two users with direct reply to a submission or a comment.  For example in Figure \ref{fig:rthread}(c), authors of comment 1 and comment 2 are involved in a dyadic interaction. Figure \ref{fig:rthread}(c) also shows examples of other dyadic interactions in the discussion thread.

\begin{table*}[t]
\centering
\resizebox{\textwidth}{!}{%
\begin{tabular}{@{}llll@{}}
\toprule
\rowcolor[HTML]{C0C0C0} 
 & Feature group (\# features) &  & Interpretation of feature values  \\ \midrule
 \multirow{-1}{*}{\rotatebox[origin=c]{90}{\textbf{Individual Factors}}} & Psychological Factors (5) & \begin{tabular}[c]{@{}l@{}}LIWC categories of \textbf{anger}, \textbf{sadness}, \textbf{anxiety}\\ VADER \textbf{positive} and \textbf{negative} sentiment\end{tabular} & {\faLongArrowUp \hspace{1pt}} values {\hspace{1pt} \faLongArrowUp \hspace{1pt}} psychological factors   \\  \cmidrule(l){2-4} 
 & Crippled Epistemology (3) & \begin{tabular}[c]{@{}l@{}}\textbf{Exclusivity in contributions:} disproportion of contributions across \\ subreddits measured through Gini coefficient\\ \textbf{Subreddit similarity to conspiracy communities:} essentially the matching criteria \resizebox{0.35cm}{!}{\includegraphics{figures/E.pdf}} \\ measured in the observation period\\ \textbf{Content similarity to conspiracy discussions:} bag of word vectors similarity to \\ top-scored discussions within the conspiracy community subreddits\end{tabular} & \begin{tabular}[c]{@{}l@{}}{\faLongArrowUp \hspace{1pt}} value {\hspace{1pt} \faLongArrowUp \hspace{1pt}} exclusivity\\ \\ {\faLongArrowDown \hspace{1pt}} value {\hspace{1pt} \faLongArrowUp \hspace{1pt}} similarity to \\ conspiracy communities\\ {\faLongArrowUp \hspace{1pt}} value {\hspace{1pt} \faLongArrowUp \hspace{1pt}} similarity to \\ conspiracy content\end{tabular}   \\ \bottomrule
& Availability (3) & \begin{tabular}[c]{@{}l@{}}\textbf{Ratio of dyadic interactions with CC:} number of dyadic interactions with CC \\ normalized by total dyadic interactions \\ \textbf{Ratio of CC in dyadic interactions:} number of CC normalized by \\ the number of all Redditors users interact with, through dyadic interactions\\ \textbf{Ratio of threads with CC:} number of discussion threads in common with CC \\ normalized by the number of all discussion threads\end{tabular} & {\faLongArrowUp \hspace{1pt}} value {\hspace{1pt} \faLongArrowUp \hspace{1pt}} availability   \\ \cmidrule(l){2-4} 
 & Information (2) & \begin{tabular}[c]{@{}l@{}}\textbf{Contribution order in dyadic interactions:} reply sent to CC is encoded as 1 \\ and reply received from CC is encoded as -1.  The feature represents the sum \\ of all such interactions converted to 1 (is sum is positive) or -1 (is sum is negative) \\ \textbf{Time lapse in dyadic interactions:} average of absolute value of time difference \\ between dyadic interactions measures in seconds\end{tabular} & \begin{tabular}[c]{@{}l@{}}{\faLongArrowDown \hspace{1pt}} value {\hspace{1pt} \faLongArrowUp \hspace{1pt}} replies received \\ from CC\\ \\ {\faLongArrowUp \hspace{1pt}} value {\hspace{1pt} \faLongArrowUp \hspace{1pt}} time lapse\end{tabular}   \\ \cmidrule(l){2-4} 
 & Reputation (2) & \begin{tabular}[c]{@{}l@{}}\textbf{Age reputation:} average of the Reddit age of the CCs users interact with \\ \textbf{Karma reputation:} average of the karma of the CCs users interact with\end{tabular} & {\faLongArrowUp \hspace{1pt}} value {\hspace{1pt} \faLongArrowUp \hspace{1pt}} reputation   \\ \cmidrule(l){2-4} 
 \multirow{-6}{*}{\rotatebox[origin=c]{90}{\textbf{Social Factors}}}  & Emotion (4) & \begin{tabular}[c]{@{}l@{}}LIWC \textbf{positive} and \textbf{negative} affect in user contributions\\ in the dyadic interactions with CC\\ \textbf{Coordination in positive and negative affect (2)}: the absolute difference between \\ the user's and the CC's positive and negative affect in dyadic interactions\end{tabular} & \begin{tabular}[c]{@{}l@{}}{\faLongArrowUp \hspace{1pt}} value {\hspace{1pt} \faLongArrowUp \hspace{1pt}} emotion in users\\ \\ {\faLongArrowDown \hspace{1pt}} value {\hspace{1pt} \faLongArrowUp \hspace{1pt}} coordination \\ in emotion with CC\end{tabular}  \\ \cmidrule(l){2-4} 
 & Group Polarization (8) & \begin{tabular}[c]{@{}l@{}}Use of \textbf{first-person singular}, \textbf{first-person plural}, the \textbf{second person} \\ and \textbf{third-person} pronouns by users in dyadic interactions with CC\\ \textbf{Coordination in the pronoun use (4)} :  the absolute difference between \\ the user's and the CC's pronoun use in dyadic interactions\end{tabular} & \begin{tabular}[c]{@{}l@{}}{\faLongArrowUp \hspace{1pt}} value {\hspace{1pt} \faLongArrowUp \hspace{1pt}} use of pronouns\\ \\ {\faLongArrowDown \hspace{1pt}} value {\hspace{1pt} \faLongArrowUp \hspace{1pt}} coordination \\ in use of pronouns with CC\end{tabular}   \\ \cmidrule(l){2-4} 
 & Self-selection (3) & \begin{tabular}[c]{@{}l@{}}\textbf{Moderated contributions:} number of moderated contributions \\ normalized by total number of contributions\\ \textbf{Negatively scoring contributions:} Total contributions with negative score \\ normalized by total contributions \\ \textbf{Contribution trend:} trend of the line fitted on number of contributions in \\ each of the six months in the observation period\end{tabular} & {\faLongArrowUp \hspace{1pt}} value {\hspace{1pt} \faLongArrowUp \hspace{1pt}} self-selection  \\ \bottomrule
\end{tabular}%
}
\caption{Table summarizing individual and social factors used in this paper. All features are written in \textbf{bold} with a concise description. A more detailed intuition behind the features is discussed in Section \ref{conspfactors}. The directionality high ({\hspace{1pt}\faLongArrowUp \hspace{1pt}}) or low ({\hspace{1pt}\faLongArrowDown \hspace{1pt}}) indicates how we should interpret the feature values and their corresponding regression coefficient signs in the logistic regression analysis (Section \ref{socimport}). 
For example, for the emotion coordination feature, low ({\hspace{1pt}\faLongArrowDown \hspace{1pt}}) value indicates high ({\hspace{1pt}\faLongArrowUp \hspace{1pt}}) coordination between users and CC, i.e., 
if FC (label 1 in regression) have high coordination with CC, the sign of beta coefficients will be negative for the emotion coordination features.}
\label{tab:summfactor}
\end{table*}

\section{Factors in Conspiratorial Engagement}
\label{conspfactors}
Table \ref{tab:summfactor} presents a concise summary of individual and social factors that are described below. 
We look at two main categories of precursors towards conspiratorial engagement---individual factors and social factors. While individual factors are designed to reflect the users' predisposition towards conspiracies, social factors capture their engagement with the members of the conspiracy communities prior to joining those communities.

\subsection{Individual Factors} 
Why do people believe in conspiracies even when there is a lack of well reasoned evidence? This question taps into a popular debate of whether conspiratorial belief emerges from psychological predisposition, or from other aspects such as an individual's exposure to biased information and to triggering events. We attempt to capture both arguments while measuring individual predisposition. 

\subsubsection{\textit{Psychological Factors: }}
We explore the presence of psychological factors through analyzing sentiment and affective words in the contributions made by FC and NC  in the observation period.  Specifically, based on previous research associating conspiratorial belief with anxiety, paranoia and insecurity \cite{butler1995psychological,goertzel1994belief}, we measure users' proclivity to such psychological factors as follows.

\begin{itemize}
    \item \textit{\textbf{Cognitive and affective processes}}:Researchers have argued that words and language reflect psychological states \cite{tausczik2010psychological}. We measure Linguistic Inquiry and Word Count (LIWC) \cite{pennebaker2001linguistic} categories of anger, sadness, and anxiety in the contributions made by users in the observation period, normalized by total number of contributions. 
    \item \textit{\textbf{Sentiment}}: We calculate average VADER sentiment scores for positive and negative sentiment \cite{gilbert2014vader} in the user's contributions during the observation period. 
\end{itemize}

\subsubsection{\textit{Crippled epistemologies: }} Sunstein et. al. coined the term to refer to a scenario when an individual's tendency to adhere to limited information sources results in their epistemological isolation \cite{sunstein2009conspiracy}. Thus, a conspiracy theory, which is otherwise unjustified relative to all the information available to the wider society, might be perfectly justified to someone whose worldview is already distorted due to the absence of relevant and ample information. The tendency to adhere to epistemologically isolated information sources increases the likelihood to accept conspiracy theories. On Reddit, users can exhibit crippled epistemologies by refraining from participating in diverse communities, participating in communities that might foster a conspiratorial worldview, and contributing content similar to the conspiratorial themes. 

\begin{itemize}
    \item \textit{\textbf{Exclusivity in contributions}}: Do the FC and NC exclusively contribute in fewer subreddits or do they spread their Reddit activity evenly over multiple subreddits? We characterize exclusivity by calculating Gini coefficient of disproportion on the subreddit contributions made by the users. The feature value varies between 0 to 1 with higher values indicating high exclusivity in subreddit contributions.
\end{itemize}

\begin{itemize}

   \item \textit{\textbf{Subreddit Similarity to conspiracy communities}}:
   Apart from subreddits in our carefully compiled list of 56 conspiracy communities, Reddit has other subreddits that even though, not dedicated to conspiracy theories, occasionally host conspiracy related discussions (for example, {\small \tt r/The\_Donald}). Higher engagement in such subreddits might indicate that users are being exposed to conspiratorial themes. The conspiracy scale introduced in section 3.2 characterizes subreddits based on how similar they are to {\small \tt r/conspiracy} compared to {\small \tt r/science}. Thus, for every user, we weigh the contributions made in each subreddit by the subreddit's score on the conspiracy scale. We consider the sum of all weighted contributions as a subreddit similarity feature. Remember that we used similar computations to match the FC and NC based on their contributions in the pre-observation period. Hence, we do not expect this feature to have significantly different values for FC and NC. However, measuring the significance of this feature in the observation period can contextualize the observations about other individual and social factors.

    \item \textit{\textbf{Content similarity to Conspiracies}}:
    Another way of measuring exposure to conspiracies is to compare the actual content produced by FC and NC with the discussions inside the conspiracy communities. Top ranking posts can distinguish subreddits along the dimensions of topics, style, audience and moderation \cite{horne2017identifying}. Hence we compile a list of top 10 scored submissions from every subreddit in the conspiracy communities as a representative corpus of conspiratorial discussions. Further, we also create a corpus for every FC and NC by combining their contributions in the observation period. Next, we create Bag of Words (BoW) representations for every corpus after cleaning the text data and removing stop words. As previously discussed, subreddits within the conspiracy communities vary in their interests (general conspiratorial discussion vs. specific conspiracies). In order to capture this variance in conspiratorial discussions, we calculate the cosine similarity scores for the user's BoW vector with all subreddits in the conspiracy communities. Finally, we take the maximum cosine similarity score as the user's content similarity feature. 
\end{itemize}

\subsection{Social Factors}
How does socializing with members of the conspiracy communities affect FC's joining behavior? We quantify the social factors by analyzing users' online interactions with  \textit{current conspiracists (CC)}. Based on Sunstein et. al's  framework \cite{sunstein2009conspiracy}, we study various statistical, temporal and linguistic aspects of the interactions between the users and CC. Below we outline the characterization of social features across various dimensions.

\subsubsection{\textbf{\textit{Availability: }}}
Conspiratorial beliefs may flourish upon \emph{availability} of conspiratorial materials \cite{sunstein2009conspiracy}. 
On Reddit, interactions with other conspiracists is what makes conspiratorial content available to users who are yet to join these communities. 
Thus, to understand the prominence of such interactions in our two user cohorts, we introduce three features.

\begin{itemize}
    \item \textbf{\textit{Ratio of dyadic interactions with CC:} } 
    Dyadic interactions are pairwise interactions (Figure \ref{fig:rthread} (c)) where either user replies to CC or vice-a-versa and can provide venues where conspiratorial content is available to users through other conspiracists. 
    We count the proportion of such dyadic interactions with CC normalized by all dyadic interactions the user has on Reddit in the observation period. 
    
    \item \textbf{\textit{Ratio of CCs in dyadic interactions:} } In addition to dyadic interactions, the amount of conspiracists engaged with users can also signal the exposure to available conspiratorial content. Do FC or NC engage with just one or multiple CC? This feature captures the number of CC that users engage with through dyadic interactions normalized by number of all Reddit users they interact with via dyadic interactions. 
    
    \item \textbf{\textit{Ratio of threads with CC}}: While dyadic interactions are  strong indicators of information exchange, users are also exposed to the contributions made by CC in the overall thread. For example, in Figure \ref{fig:rthread}, it is possible that the author of comment 4 has read comment 1 even without a direct interaction. To understand if users passively consume the content written by CC without directly engaging with them, we also consider the number of threads on which the user and CC appear together. Specifically, we calculate the user's co-presence with CC in threads by counting the total number of threads with CC normalized by the number of all threads the user participates in during the observation period. 
    
\end{itemize}

\subsubsection{\textbf{\textit{Information: }}} 
What role does information play in the conspiratorial engagement? Researchers argue that conspiratorial beliefs can be a product of informational pressure built through social interactions \cite{sunstein2009conspiracy}. For example, conspiracy theories are often initially accepted by people with low thresholds of acceptance. Informational pressure builds through social interactions with such people to the point where others even with higher acceptance threshold begin to accept the theory \cite{sunstein2009conspiracy}.   We consider CC as Redditors with lower acceptance threshold as they have already made contributions in the conspiracy communities. Towards understanding the role of information in conspiratorial engagement, we focus on two temporal characteristics of the dyadic interactions:

\begin{itemize}
    \item \textbf{\textit{Contribution order in dyadic interactions}}: Do users reply to the contributions made by CC or do they often receive a reply from CC? If the user normally replies to CC, it can indicate that she is exposed to the opinions expressed by conspiracists. Sunstein et. al. claim that this can build informational pressure that can result in conspiratorial thinking. 
    We encode every direct interaction as 1 if the user replies to CC and as -1 if the user receives as reply from CC. We aggregate this measure for all dyadic interactions by the user and consider the feature value as 1 if the sum is positive and -1 if the sum is negative. In other words, contribution sequence value of 1 indicates that the user more commonly replies to the CC. 

    \item \textbf{\textit{Time lapse in dyadic interactions}}: 
    How much time do users take to process the information they are exposed to by interacting with the CC? A small time duration between the interaction may indicate that users have less time to rationally consider all the information available and may tend to rely on other's information and judgment to form their opinion. 
    While contribution order captures whether the users contribute before or after CC, time lapse feature measures the average absolute time differences in seconds between dyadic interaction. Smaller value of time lapse means the user contributes shortly before or after CC.  
\end{itemize}

\subsubsection{\textbf{\textit{Reputation:}}}
 
When users interact with conspiracists, the reputation of conspiracists can also exert additional pressure to join the conspiratorial belief system \cite{sunstein2009conspiracy}.
Due to the reputational pressure, people often ignore their own beliefs to avoid social sanctions. We characterize reputation on Reddit with two features---account age and karma of the (CC) that NC or FC interacts with. 
\begin{itemize}
    \item\textbf{\textit{Age reputation}}: Does seniority of conspiracists exert a reputational pressure on potential joiners? We first calculate the age of a CC at the time of his last direct interaction with NC or FC in the observation period. Next, for every user in our NC, FC cohort, we calculate the age reputation feature as the average account ages of all conspiracists they engage through dyadic interactions. We consider CC's account age at the time of the latest interaction with FC in the observation period. 
    \item\textbf{\textit{Karma reputation}}: Redditors can accumulate \textit{karma} through up-votes and down-votes  on their contributions. We first find the aggregate karma of a CC user at the time of their latest direct interaction with NC or FC during the observation period. Next, for every user  in our NC, FC cohort, we calculate the average karma accumulated by all CCs that users interacted with in the observation period. 
\end{itemize}

\subsubsection{\textbf{\textit{Emotion: }}} Are emotions exchanged during interactions important towards conspiratorial belief? Sunstein et. al. argue that ``emotional selection'' could be an important aspect towards understanding the spread of conspiracies \cite{sunstein2009conspiracy}; people select content that justify their emotional state. 
Studies have also shown that discussions involving personal accounts and rumours that elicit intense emotional response are likely to spread from one person to another \cite{heath2001emotional}. Hence, we quantify this emotional snowballing by  measuring the LIWC categories of positive and negative affect words in the dyadic interactions between the user cohorts (FC and NC) and CC. 

\begin{itemize}
    \item \textbf{\textit{Affective process in dyadic interactions}}: For every direct interaction between the users and CC, we calculate the presence of LIWC's positive and negative affect category words. The aggregate positive and negative affects averaged over number of dyadic interactions represent the  affective processes in dyadic interactions. 
    \item \textbf{\textit{Coordination in affective processes}}: Do the CC reflect the same affective state as the FC and NC? While the previous feature measures the affective processes in the contributions made by FC and NC, it is also important to understand how similarly or differently the CC counteract. We measure the coordination between the affective state within dyadic interactions as follows: for every interaction, we subtract the affective state values in the contribution by CC from those in the contribution by the user. The average of this difference over all user interactions represents average coordination in the user's affective state with the CC. Lower values of the feature should indicate that users closely replicate the affective states of CC. 
    
\end{itemize}

\subsubsection{\textbf{\textit{Group Polarization: }}}
Belief in conspiracy theories is often strengthened through strong group identity \cite{Franks2013,sunstein2009conspiracy}. Prior research have found that when group members---or, in-group---have a shared sense of identity and solidarity, they often discard the arguments by outsiders---the out-group---as non-credible. This suggests that if users from our \textit{NC, FC} cohort relate to the identity of current conspiracy (CC) group members, then would likely also adopt the group's  conspiratorial beliefs. One way to measure the sense of group identity is by analyzing how users and conspiracists use pronouns in interactions \cite{job2004shared,pavalanathan2015identity,tausczik2010psychological}. For example, first person singular pronouns can signal high self and group awareness while second and third person pronouns can indicate that users are socially interactive with larger Reddit audience \cite{pavalanathan2015identity}. 

\begin{itemize}
    \item \textbf{\textit{Use of pronouns in dyadic interactions}} : We count the average use of first person singular (I, me etc.) first person plural (we, us etc.), second person and third person pronouns in the contributions made by the user in dyadic interactions with the CC
    \item \textbf{\textit{Coordination in the use of pronouns}} : We also measure the difference between the use of pronouns between the user and CC for all pronoun features mentioned above.
\end{itemize}

\subsubsection{\textbf{\textit{Self-selection: }}} Other than exposure to limited relevant information, crippled epistemology can also develop from social self-selection \cite{sunstein2009conspiracy}. As people start developing increasingly extreme conspiratorial views, they might suffer from social segregation from others with differing ideologies. 
Hence, we measure self-selection by observing the extent of social sanctions placed on a user's content contribution during their observation period. It comprises the following features.
\begin{itemize}

    \item \textbf{\textit{Moderated contributions: }} Users can feel ostracized on Reddit by having their contributions moderated.  Most subreddits have content moderation policies. Contributions that violate the subreddit rules are often removed. We calculate the number of moderated contributions normalized by total number of contributions in the observation period to understand social sanctions placed on a user's contribution.
    
    \item \textbf{\textit{Negatively scoring contributions: }} Apart from moderation, users can also face sanctions by receiving more negative scores. Contributions on a subreddit accumulate scores via upvotes and downvotes cast by others. Negative score indicates more downvotes than upvotes. Thus, we calculate contributions with total negative scores normalized by the total number of contributions. 
    
    \item \textbf{\textit{Contribution trend in the observation period: }} 
    Users may join the conspiracy communities not only because they are ostracized outside of it, but also because they generally disengage from society. To measure disengagement, we compute the decrease in their participation in the observation period. We calculate the number of contributions per month, and fit a line via least squares regression. We take the trend of this line as the contribution trend in the observation period: a negative trend corresponds to a decrease in participation.
\end{itemize}

\subsection{\textbf{Understanding the Importance of Features}}
Table \ref{tab:dists_individual} and Table \ref{tab:dists_social} in Appendix display the summary statistics and distributions for both, individual and social factors. In order to evaluate the importance of individual and social factors towards conspiratorial engagement, we construct a series of logistic regression models (see Table \ref{tab:regression}). The dependent variable is binary and represents the type of user cohort, FC (1) and NC (0). We interpret the importance of features by comparing their regression coefficients ($\beta$ values) in the logistic regression models. If features have multicollinearity---two or more features are highly related---it can lead to poor estimation of $\beta$ coefficients. 
Thus, we tested for multicollinearity in features through Variance Inflation Factor (VIF). If any feature has VIF $>$ 5.0 then the group of features is considered to have high multicollinearity \cite{sheather2009modern}. 
We found all features to have VIF $<$ 4.0  suggesting low multicollinearity. 
Additionally, all features vary in their means and standard deviations and variable types, such as counts, time in seconds and proportions. Hence we standardize the features for the regression analyses. Due to the high number of features and multiple testing, our model could have an increased risk of false significance. However, lower p-values have lesser chance of significance errors \cite{feise2002multiple}. Hence, we report p-values in different thresholds. Specifically, ($\textbf{p < 0.001}$, $\color{midp}{p < 0.01}$, $\color{lowp}{p < 0.05}$) in Table \ref{tab:regression}. Most of the p-values in the regression models are less than 0.001.

\section{Results}
\label{socimport}
How informative are social factors in predicting conspiracy joining? Previous research has already established the importance of individual predisposition in conspiratorial thinking \cite{butler1995psychological,goertzel1994belief}. Hence, we treat individual features as a baseline to ascertain the importance of social features. Specifically, we create a baseline model consisting of only individual features and then successively add six social feature groups and perform logistic regression at each step. Table \ref{tab:regression} contains the details of logistic regression performance for all models. 
In cases where new variables (features) are added to an existing model, there is a possibility of mediation effect. In other words, adding a new variable can reveal the unobserved relationship between previous independent variables and the dependent variable. Specifically, drastic changes in model coefficients ($\beta$ coefficient) along with their significance after adding new variables signal mediation. 
We observe that the significance of most of the features does not change after adding new features. This implies only limited mediation effects that won't effect our final model results. 
Additionally, the $\beta$ coefficients for features across the models are same. 
Hence, while discussing the results, we only refer to the last column of Table \ref{tab:regression}---model that contains all features. Specifically we ask four questions: 

\subsection{How important are individual features?} 
 We find that FC express more anxiety ($\beta$= 0.06), and negative sentiment ($\beta$= 0.86) compared to NC. This is consistent with qualitative studies stressing the role of negative attitudes in conspiracy adoption \cite{butler1995psychological,goertzel1994belief}. FC also show more positive sentiment compared to NC ($\beta$= 0.18). In all, FC show higher emotionality than NC. 
 Further among the \emph{crippled epistemology} features, FCs produce content more similar to the top scored discussions in the conspiracy communities ($\beta$= 0.61) and also have higher exclusivity in contributions ($\beta$= 0.09). 
 Overall, our results suggest that both negative psychological predisposition and crippled epistemology,  can inform the conspiracy joining. Our findings thus reinforce theoretical observations about the relevance of conspiratorial predisposition in the conspiracy engagement. We interpret these results further in our Discussion section.

\begin{table*}[t]
\centering
\sffamily
\renewcommand{\arraystretch}{1.1}
\resizebox{0.99\textwidth}{!}{%
\begin{tabular}{lrrrrrrr}
\hline
 & individual & + availability & + information & + reputation & +emotion & +group & +selection \\ \hline 
\textbf{INDIVIDUAL FACTORS} &  &  &  &  &  &  &  \\
\textit{\textbf{Psychological Predisposition}} &  &  &  &  &  &  &  \\
\hspace{2pt} Anger & 0.01  (0.01) & 0.01  (0.01) & 0.01  (0.01) & 0.01  (0.01) & 0.01  (0.01) & 0.01  (0.01) &  0.01  (0.01)  \\
\hspace{2pt} Anxiety &\textbf{0.06}*(0.01) & \textbf{0.06}*(0.01) & \textbf{0.06}*(0.01) & \textbf{0.06}*(0.01) & \textbf{0.07}*(0.01) & \textbf{0.07}*(0.01) & \textbf{0.06}*(0.01)   \\
\hspace{2pt} Sadness &\color{midp}{\textbf{-0.05}*(0.01)} &\color{midp}{\textbf{ -0.06}*(0.01)} & \color{midp}{\textbf{-0.06}*(0.01)} & \color{midp}{\textbf{-0.06}*(0.01)} & \color{midp}{\textbf{-0.06}*(0.01)} & \color{midp}{\textbf{-0.06}*(0.01)} & \color{midp}{\textbf{-0.06}*(0.01)}   \\
\hspace{2pt} VADER Positive Sentiment & \textbf{0.15}*(0.01) & \textbf{0.16} *(0.01) & \textbf{0.16}*(0.01) & \textbf{0.16}*(0.01) & \textbf{0.17}*(0.01) & \textbf{0.18}*(0.01) & \textbf{0.18}*(0.01)   \\
\hspace{2pt} VADER Negative Sentiment & \textbf{0.78}*(0.01) & \textbf{0.86}*(0.01) & \textbf{0.87}*(0.01) & \textbf{0.87}*(0.01) & \textbf{0.87}*(0.01) & \textbf{0.88}*(0.01) & \textbf{0.86}*(0.01)   \\
\textbf{Episteomology} &  &  &  &  &  &  &  \\
\hspace{2pt} Exclusivity in Contributions & \textbf{0.23}*(0.01) & \textbf{0.10}*(0.01) & \textbf{0.10}*(0.01) & \textbf{0.09}*(0.01) & \textbf{0.10}*(0.01) & \textbf{0.09}*(0.01) & \textbf{0.09}*(0.01)   \\
\hspace{2pt} Subreddit Similarity to conspiracy communities & \color{lowp}{\textbf{-0.07}*(0.01)} & -0.01  (0.01) & -0.01  (0.01) & -0.01  (0.01) & -0.01  (0.01) & -0.01  (0.01) & -0.01  (0.01)   \\
\hspace{2pt} Content Similarity to Conspiracies & \textbf{0.56}*(0.01) & \textbf{0.59}*(0.01) & \textbf{0.59}*(0.01) & \textbf{0.59}*(0.01) & \textbf{0.59}*(0.01) & \textbf{0.60}*(0.01) & \textbf{0.61}*(0.01)   \\ \hline
\textbf{SOCIAL FACTORS} &  &  &  &  &  &  &  \\
\textit{\textbf{Availability}} &  &  &  &  &  &  &  \\
\hspace{2pt} Ratio of Dyadic interactions with CC &  & \textbf{1.54}*(0.02) & \textbf{1.55}*(0.02) & \textbf{1.55}*(0.02) & \textbf{1.55}*(0.02) & \textbf{1.56}*(0.02) & \textbf{1.58}*(0.02)   \\
\hspace{2pt} Ratio of CC in dyadic interactions &  & \textbf{0.35}*(0.02) & \textbf{0.34}*(0.02) & \textbf{0.34}*(0.02) & \textbf{0.35}*(0.02) & \textbf{0.34}*(0.02) & \textbf{0.33}*(0.02)   \\
\hspace{2pt} Ratio of threads with CC &  & \textbf{0.1}2*(0.02) & \textbf{0.12}*(0.01) & \textbf{0.12}*(0.01) & \textbf{0.12}*(0.01) & \textbf{0.12}*(0.02) & \textbf{0.1}2*(0.02)   \\
\textbf{Information} &  &  &  &  &  &  &  \\
\hspace{2pt} Contribution Order in Dyadic Interactions &  &  & \color{midp}{\textbf{-0.13}*(0.06)} & \color{midp}{\textbf{-0.12}*(0.06)} & \color{midp}{\textbf{-0.12}*(0.06)} & \color{midp}{\textbf{-0.12}*(0.06)} & \color{midp}{\textbf{-0.13}*(0.06)}   \\
\hspace{2pt} Time Lapse in Dyadic Interactions &  &  & 0.02  (0.01) & 0.03  (0.01) & 0.02  (0.01) & 0.01  (0.01) & 0.01  (0.01)   \\
\textbf{Reputation} &  &  &  &  &  &  &  \\
\hspace{2pt} Age Reputation of CC &  &  &  & -0.02  (0.01) & -0.01  (0.01) & -0.01  (0.01) & -0.01  (0.01)   \\
\hspace{2pt} Karma Reputation of CC &  &  &  & \color{midp}{\textbf{-0.08}*(0.01)} & \color{midp}{\textbf{-0.08}*(0.01)} & \color{midp}{\textbf{-0.08}*(0.01)} & \color{midp}{\textbf{-0.08}*(0.01)}   \\
\textbf{Emotion} &  &  &  &  &  &  &  \\
\hspace{2pt} LIWC Positive affect &  &  &  &  & 0.02  (0.01) & 0.01  (0.01) & 0.01  (0.01)   \\
\hspace{2pt} Coordination in positive affect &  &  &  &  & \textbf{-0.05}*(0.01) &\textbf{-0.05}*(0.01) &\textbf{-0.05}*(0.01)   \\
\hspace{2pt} LIWC Negative affect &  &  &  &  & 0.03  (0.01) & 0.04  (0.01) & 0.03  (0.01)   \\
\hspace{2pt} Coordination in negative affect &  &  &  &  & \textbf{-0.01}*(0.01) & \textbf{-0.01}*(0.01) & \textbf{-0.02}*(0.01)   \\
\textbf{Group Polarization} &  &  &  &  &  &  &  \\
\hspace{2pt} First person singular &  &  &  &  &  & \textbf{0.04}*(0.01) & \textbf{0.04}*(0.01)   \\
\hspace{2pt} Coordination in first person singular &  &  &  &  &  & 0.01  (0.01) & 0.01  (0.01)   \\
\hspace{2pt} First person plural &  &  &  &  &  & \textbf{0.04}*(0.01) & \textbf{0.04}*(0.01)   \\
\hspace{2pt} Coordination in first person plural &  &  &  &  &  & \textbf{-0.04}*(0.01) & \textbf{-0.04}*(0.01)  \\
\hspace{2pt} Second person &  &  &  &  &  & 0.01  (0.01) & 0.01  (0.01)   \\
\hspace{2pt} Coordination in second person &  &  &  &  &  & 0.02  (0.01) & 0.01  (0.01)   \\
\hspace{2pt} Third person &  &  &  &  &  & 0.01 (0.01) & 0.01 (0.01)   \\
\hspace{2pt} Coordination in third person &  &  &  &  &  & \textbf{-0.0}5*(0.01) & \textbf{-0.05}*(0.01)   \\
\textbf{Self-selection} &  &  &  &  &  &  &  \\
\hspace{2pt} Moderated contributions &  &  &  &  &  &  & \textbf{0.06}*(0.01)     \\
\hspace{2pt} Negatively scoring contributions &  &  &  &  &  &  & \textbf{0.18}*(0.01)     \\
\hspace{2pt} Contribution trend in the observation period &  &  &  &  &  &  & \textbf{0.09}*(0.01)     \\ \hline
intercept & -0.10 & 0.07 & 0.07 & 0.07 & 0.07 & 0.06 & 0.07 \\
Adjusted $R^{2}$ & 0.12 & 0.21 & 0.22 & 0.23 & 0.23 & 0.24 & 0.26  \\
Accuracy & 0.64 & 0.71 & 0.72 & 0.72 & 0.72 & 0.72 & 0.73 \\
Precision & 0.63 & 0.72 & 0.73 & 0.73 & 0.73 & 0.73 & 0.74 \\
Recall & 0.65 & 0.68 & 0.68 & 0.69 & 0.69 & 0.69 & 0.70 \\ \hline
\end{tabular}%
}
\caption{Results of the regression analysis. In each column we successively add social feature groups to the individual features and observe $\beta$ values and adjusted $R^{2}$. Significant $\beta$ values are followed by *. We color code the numbers according to p-values as follows: ($\textbf{p < 0.001}$, $\color{midp}{p < 0.01}$, $\color{lowp}{p < 0.05}$). $\beta$ values stay fairly consistent throughout all models and most features do not change their significance indicating limited or partial mediation effects. Accuracy, precision and recall were calculated using five-fold cross-validation. Our $R^{2}$ values are fairly consistent with the ranges reported by other researchers studying complex social phenomena \cite{chang2020don,wang2016modeling}. Note that various social feature groups can be successively added to the individual feature model in any order. Hence, we do not use regression performance from this table to compare social feature groups with each other.}
\label{tab:regression}
\end{table*}

\subsection{How important are social features?} 

We treat individual features' model as a baseline to compare how much value do social features add to the regression model. Below, we discuss each social feature group separately.

\subsubsection{\textit{\textbf{Availability}}:} 
By adding \textit{availability} features we observe improvement in the pseudo $R^{2}$ compared to the individual factors model (0.21 vs. 0.12). That is, the model containing both individual and availability features fits better than the model with just individual features. 
Specifically, in comparison to NCs, FC users have higher proportion of dyadic interactions ($\beta = 1.58$) and more threads in common ($\beta$= 0.12) with CCs. 
Moreover, out of all Redditors they interact with, the proportion of CCs they interact with, is larger for FC ($\beta$= 0.33) compared to NC.  Upon adding the availability features, the subreddit similarity in individual factors becomes insignificant, indicating a possibility of partial mediation effect. 
Overall, high number of dyadic interactions with CC and co-presence with CC indicate intimacy with current conspiracists; intimacy is one of the four tie strength dimensions proposed by Granovetter \cite{Granovetter1973TheTies}. This may suggest that FC form strong ties with CC in the observation period. 


\subsubsection{\textit{\textbf{Information}}: } Information features capture the order of contributions, i.e, whether users usually reply to the CC or vice a versa, and the time lapse between the dyadic interaction. 
Remember that negative feature value of contribution order indicates more replies received from CC as opposed to positive which indicates more replies sent to CC. 
The negative and significant beta coefficient for the contribution order feature ($\beta$= -0.13) thus implies that FC often receive more direct replies from the current conspiracists. Together with increased direct interactions with CC in general, this might reflect efforts on part of CC to engage with FC. We find no significant differences in the time lapse between the dyadic interactions of user cohorts and CC.

\subsubsection{\textit{\textbf{Reputation}: }} While there is no difference between the seniority of CC that FC and NC interact with, they do differ in the average karma. On Reddit, comment and submission karma can indicate how well the user's opinions are accepted by other Reddit users. We find that FC interact with CC having lower karma ($\beta$=-0.08). It is possible that the current cospiracists who feel rejected by other Reddit users through lower karma are reaching out to a group of users predisposed towards conspiratorial thinking---FC---by direct interactions outside of the conspiracy communities.

\subsubsection{\textit{\textbf{Emotions}: }} What role do emotions in interactions play in conspiratorial engagement? Adding \textit{emotion} features, we observe an interesting coordination between affective state of FC and NC. Firstly, FC and NC do not differ in the positive and negative affect word use in dyadic interactions with CC. However, CC still closely reflect the affective state of FC more compared to NC. To elaborate, coordination features are calculated by taking the absolute difference between the user's and CC's respective emotion states. Hence, lower values represent higher coordination. Thus, FC have higher coordination of negative ($\beta$=-0.02) and positive ($\beta$=-0.05) affective state with CC  compared to NC.

\subsubsection{\textit{\textbf{Group polarization}: }} For the group polarization features, we measure the use of pronouns by the users in dyadic interactions and how similar their pronounce usage is to the use of pronouns by CC in dyadic interactions. While talking with CC, FC use  first person pronouns more compared to NC. Interestingly, FC also have higher coordination in the first person plural ($\beta$=-0.04) and third person ($\beta$=-0.05) use with CCs. Previous research has associated higher use of first person plural and third person pronouns with strong group identity \cite{inigo2004use,tausczik2010psychological}. This means that even before joining conspiracy communities, FC communicate in the language of ``we'', ``us'', ``ours'' with CC expressing higher group identity.

\subsubsection{\textit{\textbf{Selection}: }} How does self-selection affect conspiracy engagement? Do users get more sanctions and negative feedback from others? 
\textit{Selection} features capture the amount of moderation, negative scores and users' disengagement during the observation period. We find that all \textit{selection} features are significant. Specifically, FC have more moderated ($\beta$=0.06) and negatively scored contributions ($\beta$=0.18) compared to NC. However, despite facing more social sanctions, FC still increase their contribution rate compared to NC in the observation period ($\beta$=0.09).

\begin{figure*}[t]
    \centering
        \includegraphics[width=0.45\textwidth]{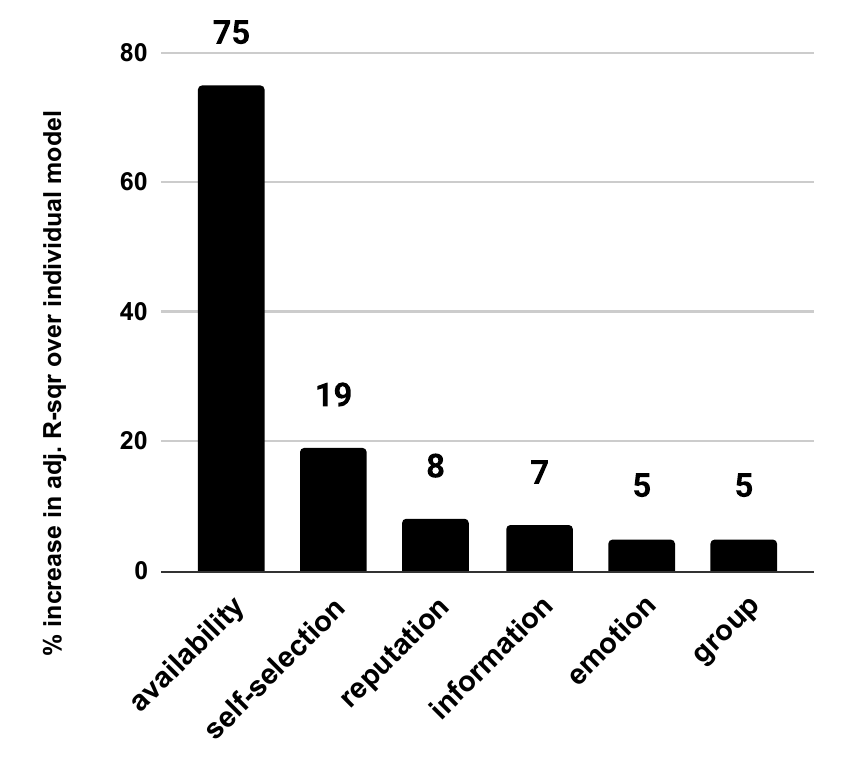}

    ~
    \caption{(a) Six social feature groups and their relative importance as percent increase in $R^{2}$ compared to individual features (more details in Section \ref{reldocimport}). Note that we add only one social feature group at a time to the individual model and calculate the percent increase in the $R^{2}$. For example, availability features improve the $R^{2}$ of individual model by 75\% while self-selection feature improve it by 19\%. }
    \label{fig:social_importance}
\end{figure*}

\subsection{How important is each of the social feature groups?}
\label{reldocimport}

All social feature groups contain at least one significant predictor of joining conspiracy communities. In fact, some social factors are overall better predictors than individual factors. Notably, the ratio of dyadic interactions with CC ($\beta$=1.58) is the single most important feature in the regression model. Furthermore, iteratively adding each group of social features consistently increases model performance. These findings corroborate that social factors play a sizable role in predicting users joining conspiracy communities \textit{even after controlling for individual factors}.  To understand the relative importance of feature groups amongst social factors, we add each of them separately to the individual features, and compare their percent increase in explained variance ($R^2$). Figure \ref{fig:social_importance} displays a bar chart indicating relative increase in the $R^{2}$ value over individual features.
We find that among all the social features, \textit{availability} features are the most informative (75\% increase), followed by \textit{selection} (19\%), \textit{reputation} (8\%), \textit{information} (7\%), \textit{emotion} (5\%), and \textit{group polarization} (5\%). 
In summary, we find evidence 
that the different social factors hypothesized in \cite{sunstein2009conspiracy} capture specific, complementary, and relevant aspects of the joining behavior---although in varying amounts.

\begin{table*}
\centering
\resizebox{\textwidth}{!}{%
\begin{tabular}{@{}l
>{\columncolor[HTML]{C0C0C0}}l ccc@{}}
\cmidrule(l){2-5}
 & \multicolumn{1}{c}{\cellcolor[HTML]{C0C0C0}} & \cellcolor[HTML]{C0C0C0} & \cellcolor[HTML]{C0C0C0} & \cellcolor[HTML]{C0C0C0} \\
\multirow{-2}{*}{} & \multicolumn{1}{c}{\multirow{-2}{*}{\cellcolor[HTML]{C0C0C0}Features}} & \multirow{-2}{*}{\cellcolor[HTML]{C0C0C0}\begin{tabular}[c]{@{}c@{}}Test 1 Accuracy\\ (20\% remaining)\end{tabular}} & \multirow{-2}{*}{\cellcolor[HTML]{C0C0C0}\begin{tabular}[c]{@{}c@{}}Test 2 Accuracy\\ (10\% Generalist Population)\end{tabular}} & \multirow{-2}{*}{\cellcolor[HTML]{C0C0C0}\begin{tabular}[c]{@{}c@{}}Test 3 Accuracy\\ (10\% Specialist Population)\end{tabular}} \\ \cmidrule(l){2-5} 
\multicolumn{1}{c}{} & individual & 0.64 & 0.66 & 0.62 \\
\multicolumn{1}{c}{\multirow{-2}{*}{\begin{tabular}[c]{@{}c@{}}Training\\ (80\% whole population)\end{tabular}}} & individual+social & 0.73 & 0.71 & 0.74
\end{tabular}%
}
\caption{Full model tested on the generalist and specialist population. The experiment was repeated five times. The accuracy values are the average of 5 experiments. We find that accuracy values are similar across three tests and increase after adding social factors. This indicates that social factors are informative in predicting conspiracy joining regardless of the topic of discussion in the conspiracy subreddit. }
\label{tab:new_gsval}
\end{table*}

\subsection{Are social factors similarly informative for topic-specific vs. general conspiracy joining? }  
\label{robust}
Conspiracy communities are internally diverse, including subreddits discussing conspiracy theories in \textit{general}, like {\small \tt r/conspiracy}, along with ones with a narrow focus on \textit{specific} theories, like advanced energy weapons ({\small \tt r/TargetedEnergyWeapons}) and alien encounters ({\small \tt r/reptilians}). 
Previous researchers have characterized Reddit users as generalists and specialists, and found significant behavioral differences \cite{waller2019generalists}. In particular, generalists engage with more diverse sets of subreddits than specialists do.\footnote{We adapt the definition of generalist and specialist to the context of conspiracies. We characterize the scope of the conspiracies discussed in the most frequented subreddit, rather than the user's activity diversity across subreddits \cite{waller2019generalists}.} 
We test the robustness of social factors as predictors towards engagement with conspiracy communities, and study their informativeness for users who focus on general vs. specific conspiracy communities. 
We first labeled every subreddit in the conspiracy communities as ``general'' or ``specific''. All subreddits in the conspiracy communities had descriptions that allowed us to identify, with confidence, whether the subreddit was topic-specific or not. For example, {\small \tt r/HOLLOWEARTH} is a topic-specific subreddit that purports that planet Earth is internally hollow; it self describes as a subreddit ``[...] for celebrating and sharing the knowledge of our hollow earth''. Whereas, the general subreddit {\small \tt r/conspiro} ``[...] allow[s] intelligent discussion on any topic''. 
Next, we classify every FC user as a ``generalist'' or a ``specialist'' based on the subreddit they have the highest contributions since joining the conspiracy communities. In other words, if their future conspiracy community subreddit of highest contribution was ``specific'', then the FC was labeled as ``specialist'', otherwise ``generalist''.

To understand how well individual and social factors can predict different types of FC users---generalists and specialists---we performed a series of tests. First, we trained models on two sets of features: (1) only individual and (2) individual plus social factors (corresponds to the first and the last columns in Table \ref{tab:regression}, respectively). We calculated the accuracy of the models on a randomly sampled held out set of 10\% of generalist FC and 10\% of specialist FC (along with their matched NC). We repeated this experiment five times and reported the average accuracy results in Table \ref{tab:new_gsval}. In particular, we measured if the models performed equally well on the two cohorts of generalist and specialist FC users, and if social factors improved model performance in both cases. 

Indeed, accuracy values do not change significantly based on the test population. For example, in the model with only individual features, accuracy on a random sample of the population is 0.64, whereas accuracy for generalists and specialists only is respectively 0.66 and 0.62. This indicates that model performance is comparable for generalist and specialist users. Furthermore, we find that adding social factors increases model accuracy in all test conditions (0.73 for random test, 0.71 for generalists and 0.74 for specialists). In other words, conspiratorial joining can be predicted from social factors regardless of how narrow the topic of the subreddit.

\section{Discussion}
Our current understanding of social factors in conspiracy adoption is assembled from mainly theoretical studies. This work calls into attention the importance of \textit{empirically} studying how interactions with current members can influence the user's joining into the conspiracy communities. By proposing a theoretical-motivated, quantitative operationalization of social factors across six dimensions, we take a step in this direction.
Specifically,  we provide empirical representation of social features proposed by \citeauthor{sunstein2009conspiracy} in six groups: 1) importance of social availability of conspiracists, 2) informational pressure, 3) reputational pressure, 4) emotional snowballing, 5) group identity, and 6) self-selection towards conspiracy adoption. We compared the social factors with the strong baseline of individual factors from literature \cite{butler1995psychological,goertzel1994belief,Darwin2011BeliefSchizotypy,Sunstein2009ConspiracyCures}, and found that social factors are crucial precursors of joining conspiracy communities. Not only social factors as a whole add significant explanatory power over individual factors, each of the six dimensions contains significant predictors that capture separate and complementary facets of conspiracy theory adoption. Our findings bring forth several implications for understanding how the conspiracy communities form, how they maintain their echo chamber, and how social exclusion may lead to joining \footnote{while referencing the $\beta$ coefficients downstream, we refer to the numbers only from the last column of Table \ref{tab:regression}}.

\subsection{Selection, Evocation, and Manipulation Among Joiners}
How do conspiracy communities grow? 
We refer to \citeauthor{buss1987selection}'s proposed causal mechanisms of individual---community correspondence to understand how {future conspiracists} first select, and then assimilate into conspiracy communities \cite{buss1987selection}. Buss presents three key mechanisms: \textit{selection}, whereby individuals decide to participate in a social group based on personal preference or mere proximity; \textit{evocation}, where individuals elicit emotional responses from the group in order to make connections; and \textit{manipulation}, whereby they use their position in newly found environment to change it. 
We find that users produce content similar to discussions within the conspiracy communities ($\beta$= 0.61) and increasingly interact with  conspiracy members before joining the conspiracy communities ($\beta$ = 1.58). Together, these findings show the hallmarks of the \textit{selection} process of conspiracists' future social group. Next, we observe \textit{evocation} in how future and current conspiracists coordinate their online messages. Specifically, we find that the affective states and group identity signals of future conspiracists closely mirror those of their future social group---current conspiracists (See Table \ref{tab:regression} last column). 
The present work studies the precursors to joining the social group, and therefore it does not directly observe future conspiracists' behavior after they become members of the community. According to Buss, in the \textit{manipulation} phase, conspiracists would take on an active role in gatekeeping their newly found community. 
In particular, we see that current conspiracist may play a crucial role in recruiting new members though dyadic interactions ($\beta$ = 1.58), although whether that is on purpose remains unanswered. Dyadic interactions between future and current conspiracists are at the nexus of the former's selection of a community to belong, and the latter's attempts to shape it. Studying this negotiation is essential to understand how conspiracy communities self-sustain and thrive. Our work  offers crucial insight in this direction.

\subsection{Conspiracy Communities as an Echo Chamber}
Previous studies posit that consumers of conspiracy-like content are likely to aggregate in homophilic clusters---i.e. ``echo chambers'' \cite{bessi2015science}. In fact, conspiracy theorists are renowned for their commitment to conspiratorial attitudes, and this may come from limited access to contradicting information early on \cite{sunstein2009conspiracy}. 
Our results empirically corroborate previous work's hypothesis that future conspiracists live in an information bubble. In fact, not only do they contribute content similar to conspiracy discussions ($\beta$ = 0.61), they also engage disproportionately in subreddits similar to those in the conspiracy communities ($\beta$ = 0.09). Apart from such informational isolation, echo chambers can also result from fragmentation of communities where like minded people come together to discuss ideas through a very narrow world-view. Our results indicate that along with exposure to conspiratorial material, users also directly interact with members of the conspiracy communities ($\beta$ = 1.58) and current conspiracists make up a significant fraction of their social circle on Reddit ($\beta$ = 0.33). While we do not claim that the information discussed in such interactions is strictly conspiracy related, the relevance of both epistemological and social isolation indicate that  future conspiracists may be living in their own informational echo chamber circulating similar conspiratorial content even prior to joining conspiracy communities.

\subsection{Social Stigma and Joining the Conspiracy}
We uncover an important factor in joining conspiracy communities: marginalization from other communities. Through self-selection features we find that future conspiracists are ostracized from subreddits outside of the conspiracy communities through negative feedback from other members of those communities ($\beta$ = 0.18) and content moderation ($\beta$ =0.06), significantly more than non-conspiracists. Future conspiracists express anxiety and negative sentiment in the months leading up to their joining (psychological predisposition, Table \ref{tab:regression}). We give an interpretation of how this may affect the formation of conspiracy groups. While discussing deviance as a social construct, Becker proposed that groups create rules to define what they (subjectively) consider to be desirable behavior \cite{becker2008outsiders}. As a consequence, people who break such rules are labeled as deviants and criminalized. Despite their popularity, the public image of conspiracies is still tainted,  and conspiratorial thinking bears the stigma of deviance. A two-fold process can then explain joining conspiracy communities. First, social sanctions make users feel like outsiders in mainstream subreddits. Such socially outcast users then find home in the conspiracy communities for their rejected thoughts.

\subsection{Implications}

\subsubsection{Implications for content moderation: } Researchers have found that moderators notice repeat offenders---users who have already faced sanctions before---and partially focus their moderation efforts on them \cite{lampe2004slash}. We observe that future conspiracists already start facing social sanctions in terms of content moderation ($\beta$ = 0.06) and negative karma ($\beta$ = 0.18) prior to joining conspiracy communities. We argue that this type of ostracizing may exacerbate the segregation of future conspiracists and drive them to contribute in communities that accept their conspiratorial worldview. Therefore, community managers and social platform may play a determining role in the creation of conspiracy communities. A mindless application of norms that are too rigid may ultimately ostracize non-conforming individuals, thus running the risk of driving them into fringe groups.

\subsubsection{Theoretical Implications: }  Our study engages with methodological challenges of using observational data to implement a theoretical framework for understanding social factors in conspiratorial joining.  We explore beyond the purely theoretical framework and quantitatively establish the importance of different social factors on large-scale online discussion communities. We further test the generalizability of our individual and social factors for topic-specific and general conspiracy joining. Although prior work largely framed conspiracism as an individual pursuit, focusing on psychological disposition \cite{goertzel1994belief,butler1995psychological,hofstadter2012paranoid} and epistemological characteristics \cite{Sunstein2009ConspiracyCures} our results support a socio-constructionist view of conspiracy theory. In particular, this view grants drawing the parallel between discussing conspiracy theories, and entering the community that hosts those discussions. Our analysis and results focusing on social factors in conspiracy engagement provide us with a unique opportunity to consider conspiracies as social movements---``a network of interactions between groups of individuals or organizations, engaged in a political or cultural conflict, on the basis of a shared collective identity'' \cite{diani1992concept}.  For instance, in the case of conspiracist belief, collective identity based on political ideology can lead to upholding different types of anti-government conspiracies. Republicans are more likely to believe that a ``Deep State'' is colluding against President Trump \cite{Newpollt65online} whereas Democrats more commonly believe that 9/11 was an inside job \cite{Morethan65online}. 
Characterizing conspiracies as social movements becomes more relevant when conspiracism has a potential to turn into  conspiracy \textit{activism} towards a cause with detrimental consequences. Consider the anti-vaccination movement set in motion by anti-vaccination conspiracies which has directly resulted in lower herd immunity. Analyzing conspiracies through a social movement lens can open up further research avenues exploring how conspiracists frame their narrative, mobilize informational resources, and ultimately coordinate collective action.

\section{Limitations}
Our work has some limitations which also pave the way for promising future directions. We characterize engagement within the conspiracy communities based on the number of contributions a user makes in conspiracy subreddits; contribution based approach is a common methodological choice made when studying users in social media \cite{Hamilton2017LoyaltyCommunities}. A more robust definition of engagement could involve analyzing the topics and synchronicity of the user's contribution content with that of the community.  For example, the criteria for selecting FC could be made stricter by keeping only those who discuss topics similar to conspiracies. 
Additionally, similar to any observational quantitative research, we can not infer true causality. While acknowledging this, we believe that our work is an important step towards understanding the variety of statistical, temporal and linguistic social factors towards conspiracy joining in a complex, real world setting. However, we take this opportunity to invite further qualitative studies investigating conspiracy joining using insights provided in our work. While testing the robustness of social features we consider only one dichotomy---topic specific and general conspiracy discussion subreddits. Fruitful path for future exploration could be to check how social factors vary for conspiracy joining in smaller or larger subreddits or, political or non-political conspiracy subreddits. Finally, our results exemplify conspiracy joining on just one online platform---Reddit. We do not know how these results translate to other platforms such as Facebook or Gab with various levels of content moderation. We encourage future researchers to build up on our findings and to explore conspiracy joining across multiple platforms.

\section{Conclusion}
Currently, our understanding of social factors in conspiracy adoption is patched together by mainly theoretical and very few empirical studies. This work calls into attention the importance of systematically studying how interactions with conspiracists can influence the user's joining into the conspiracy communities. Using a theory driven framework of social factors across six dimensions, we perform a retrospective case control study of \textit{future conspiracists} and compare them with \textit{non conspiracists}. We not only find that the social factors are important but that conspiracy joining can be explained at least partially by at least one feature in each group. 
Given these findings, we offer a unique, empirically backed perspective on the life-cycle of conspiracists, echo chambers in conspiracy communities and the effect of social exclusion in conspiracy engagement.

\section{Acknowledgments}
This paper would not be possible without the valuable feedback from the entire Social Computing lab group at Virginia Tech and the University of Washington. This work is partially supported by an ICTAS Junior Faculty award and NSF grant IIS-1755547.


\bibliographystyle{ACM-Reference-Format}
\bibliography{ref,samory1,new}

\appendix

\newpage
\section{Appendix}
\begin{table*}[h]

\centering
\sffamily \scriptsize
\begin{tabular}{@{}llllll@{}}
\toprule
\textbf{INDIVIDUAL FACTORS} & \textbf{min} & \textbf{max} & \textbf{mean} & \textbf{std} & \textbf{distribution} \\ \midrule
\textbf{Psychological Predisposition} &  &  &  &  &  \\
Anger & 0.0 & 0.15 & 0.07 & 0.01 & \parbox[c]{0.8em}{\includegraphics[width=0.5in]{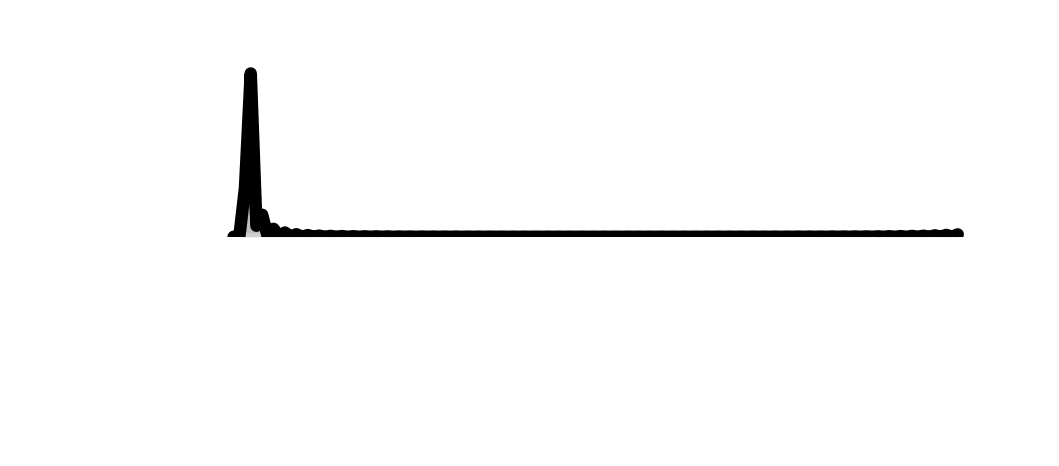}}  \\
Anxiety & 0.0 & 0.36 & 0.04 & 0.04 & \parbox[c]{0.8em}{\includegraphics[width=0.5in]{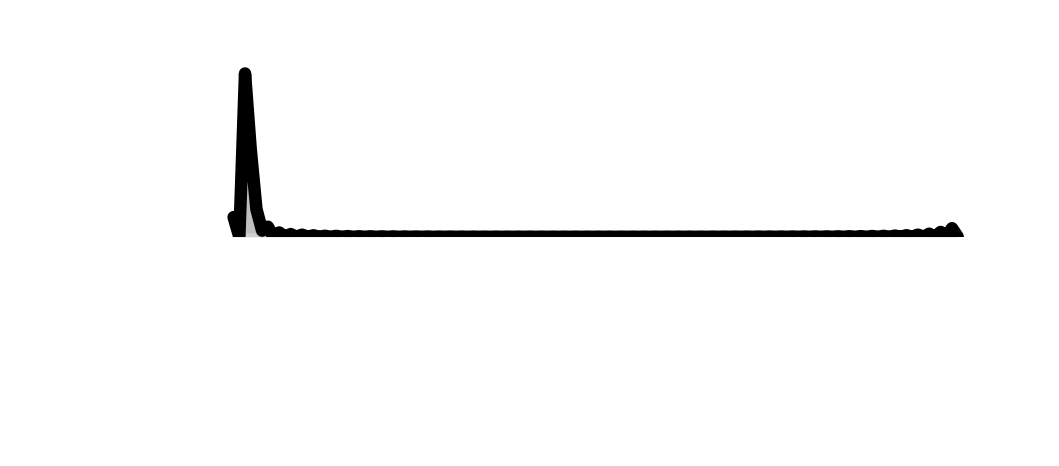}}  \\
Sadness & 0.0 & 0.20 & 0.04 & 0.03 & \parbox[c]{0.8em}{\includegraphics[width=0.5in]{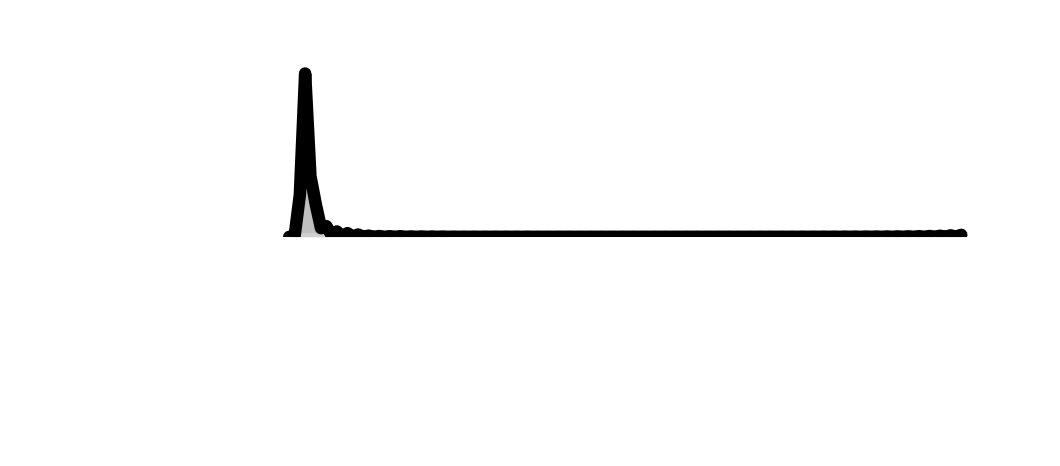}} \\
VADER Positive Sentiment & 0.0 & 1.00 & 0.14 & 0.04 & \parbox[c]{0.8em}{\includegraphics[width=0.5in]{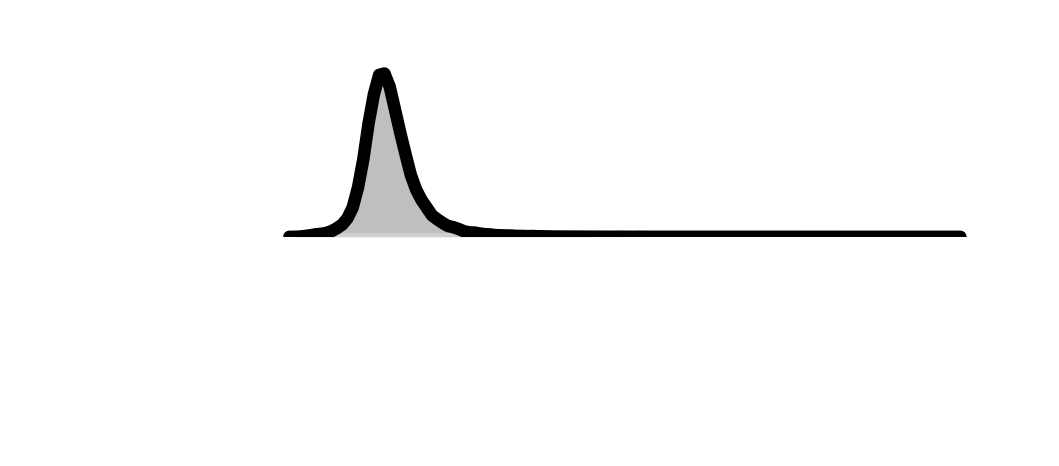}} \\
VADER Negative Sentiment & 0.0 & 0.78 & 0.08 & 0.03 & \parbox[c]{0.8em}{\includegraphics[width=0.5in]{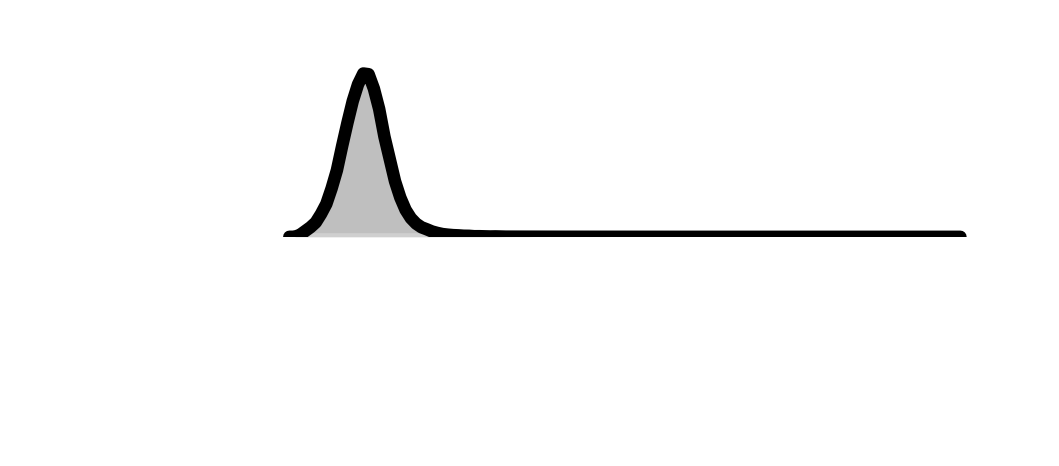}} \\
\textbf{Episteomology} &  &  &  &  &  \\
Exclusivity in Contributions & $\sim$0 & 0.97 & 0.58 & 0.18 & \parbox[c]{0.8em}{\includegraphics[width=0.5in]{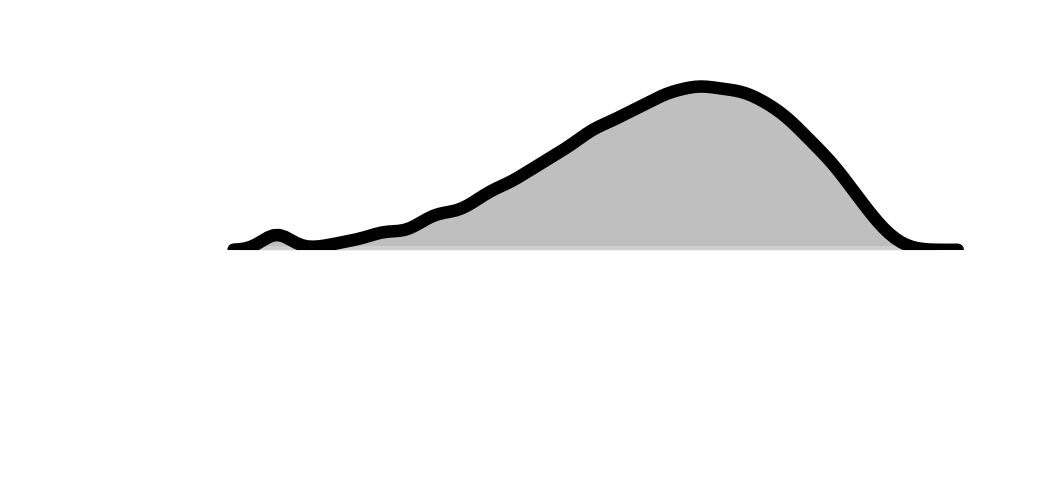}} \\
Subreddit Similarity to conspiracy community & -0.32 & 27.3 & 1.88 & 3.67 & \parbox[c]{0.8em}{\includegraphics[width=0.5in]{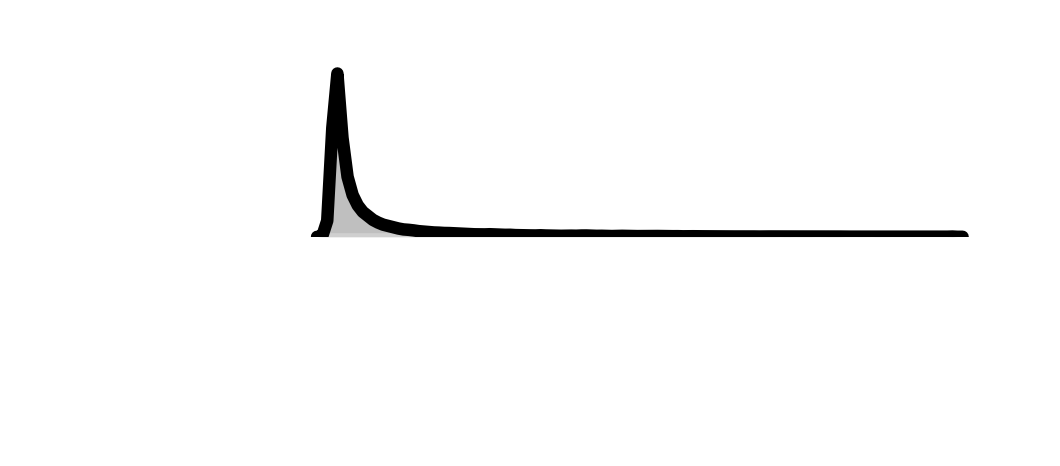}} \\
Content Similarity to Conspiracies & $\sim$0 & 0.76 & 0.41 & 0.13 & \parbox[c]{0.8em}{\includegraphics[width=0.5in]{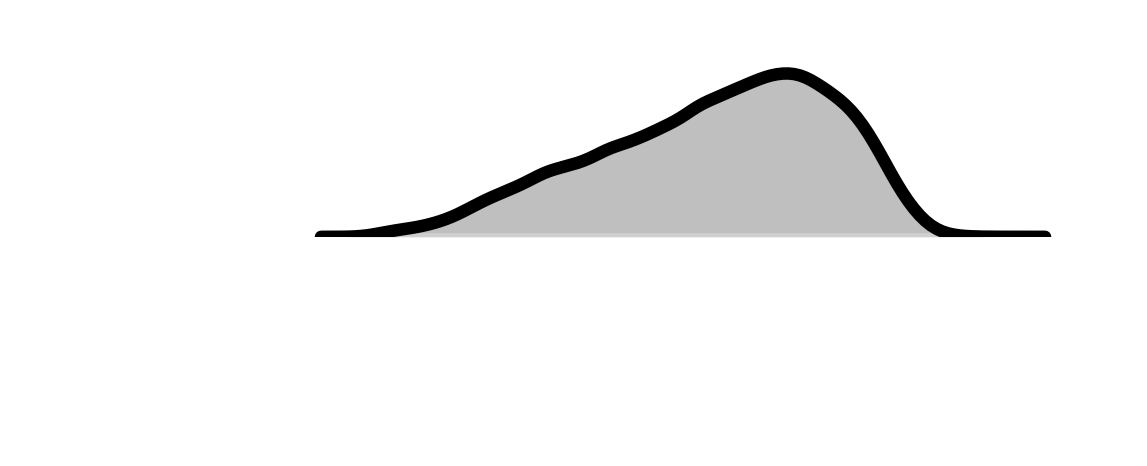}} \\ \bottomrule
\end{tabular}%
\caption{Descriptive statistics and distribution plots for individual factors. }
\label{tab:dists_individual}
\end{table*}

\begin{table}

\centering
\sffamily \scriptsize
\begin{tabular}{@{}llllll@{}}
\toprule
\textbf{Social Factors} & \textbf{min} & \textbf{max} & \textbf{mean} & \textbf{std} & \textbf{distribution} \\ \midrule
\textbf{Availability} &  &  &  &  &  \\
Ratio of Dyadic interactions with CC & 0.0 & 1.0 & 0.06 & 0.05 & \parbox[c]{0.8em}{\includegraphics[width=0.5in]{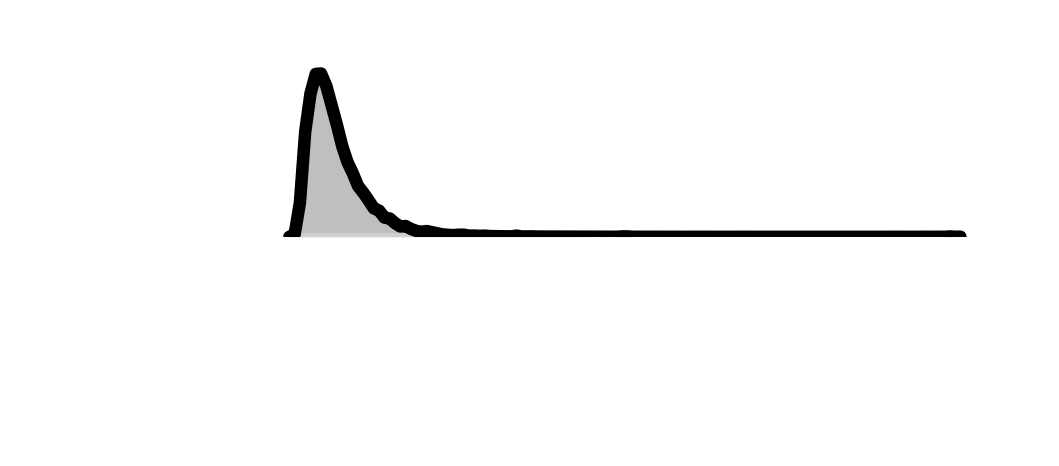}} \\
Ratio of CC in dyadic interactions & 0.0 & 1.0 & 0.09 & 0.06 & \parbox[c]{0.8em}{\includegraphics[width=0.5in]{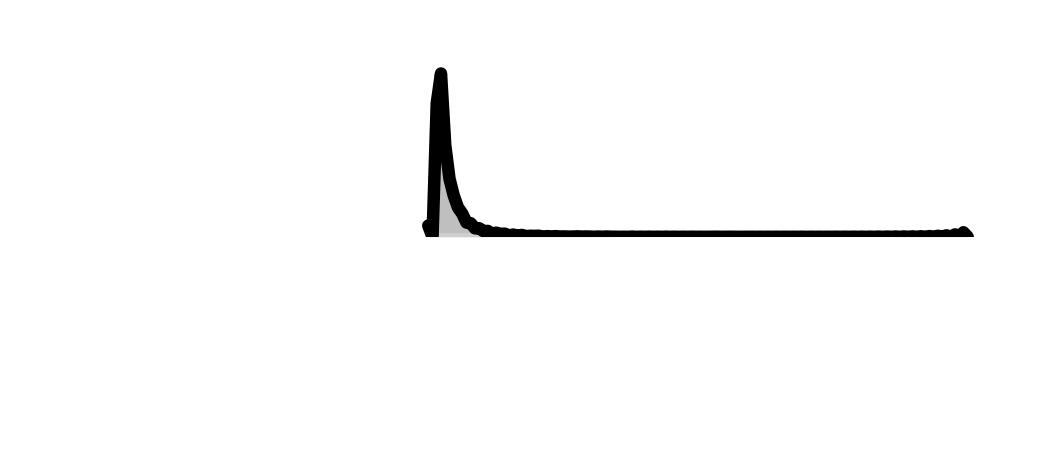}}  \\
Ratio of threads with CC & 0.0 & 1.0 & 0.10 & 0.08 & \parbox[c]{0.8em}{\includegraphics[width=0.5in]{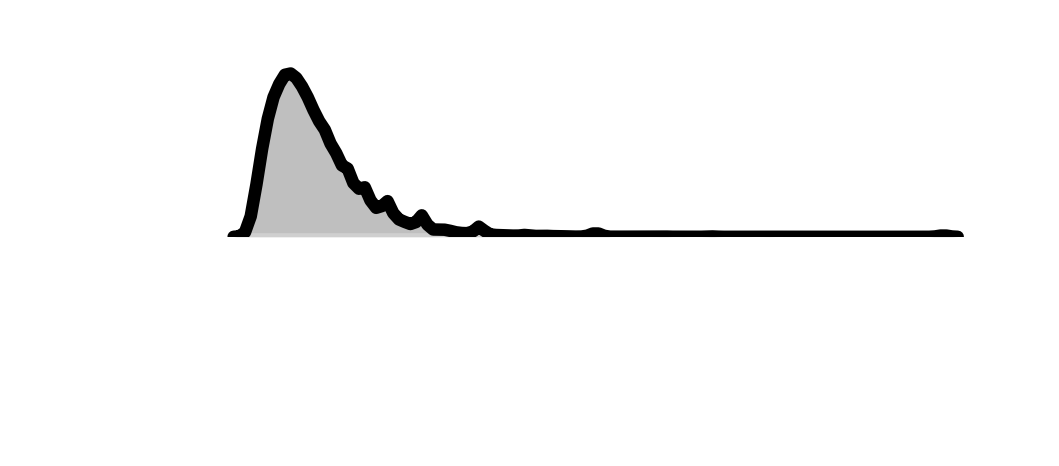}} \\
\textbf{Information} &  &  &  &  &  \\
Contribution Order in Dyadic Interactions & -2868 & 787 & -0.66 & 118 & \parbox[c]{0.8em}{\includegraphics[width=0.5in]{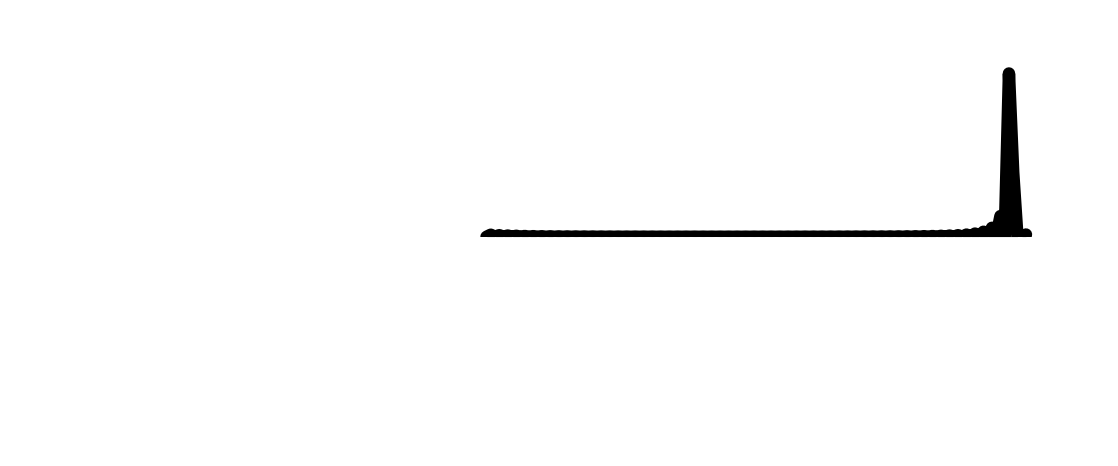}} \\
Time Lapse in Dyadic Interactions (seconds) & 16.0 & 30716740 & 50321 & 277346 & \parbox[c]{0.8em}{\includegraphics[width=0.5in]{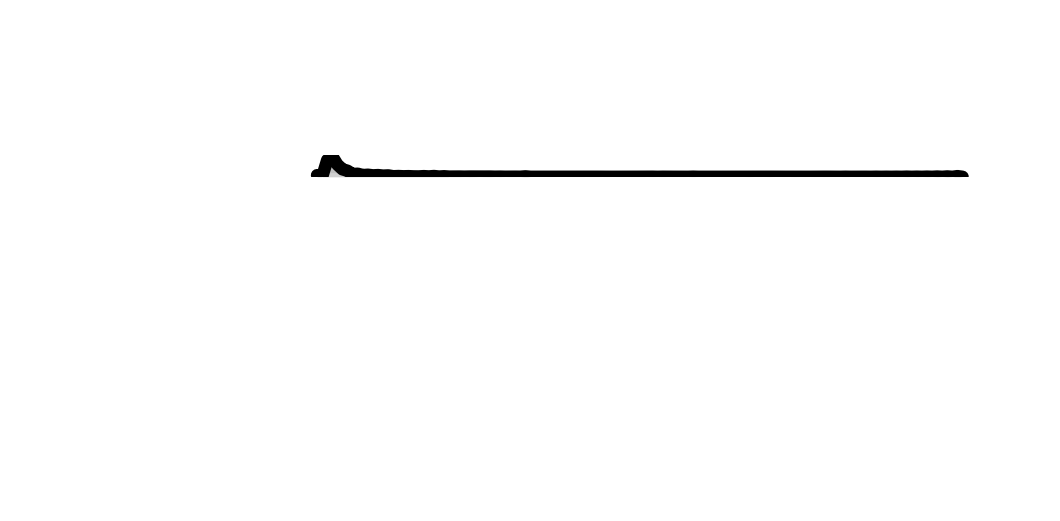}} \\
\textbf{Reputation} &  &  &  &  &  \\
Age Reputation of CC (months) & 0.03 & 113.0 & 22.88 & 10.73 & \parbox[c]{0.8em}{\includegraphics[width=0.5in]{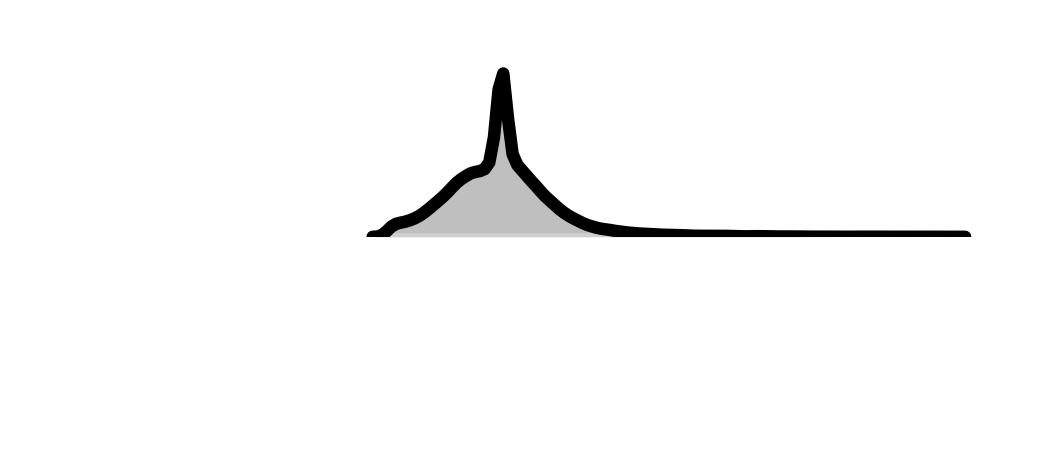}} \\
Karma Reputation of CC & -12090 & 12015457 & 189715 & 363908 & \parbox[c]{0.8em}{\includegraphics[width=0.5in]{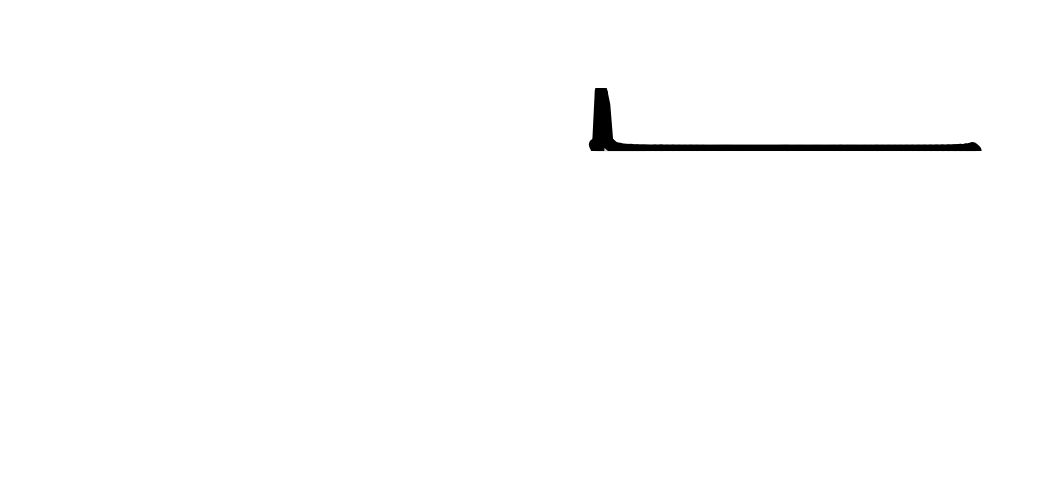}} \\
\textbf{Emotion} &  &  &  &  &  \\
LIWC Positive affect & 0.0 & 0.60 & 0.12 & 0.10 & \parbox[c]{0.8em}{\includegraphics[width=0.5in]{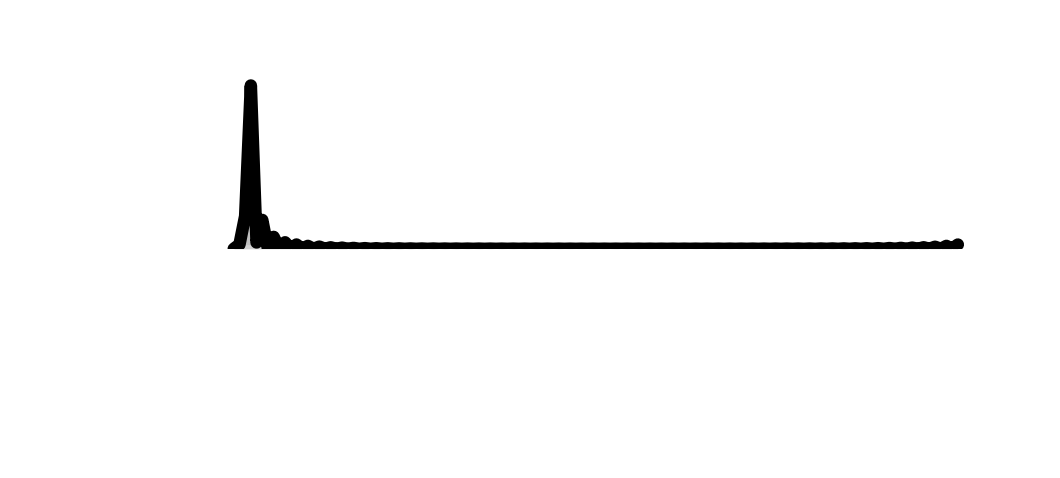}}  \\
Coordination in positive affect & 0.0 & 0.80 & 0.18 & 0.14 & \parbox[c]{0.8em}{\includegraphics[width=0.5in]{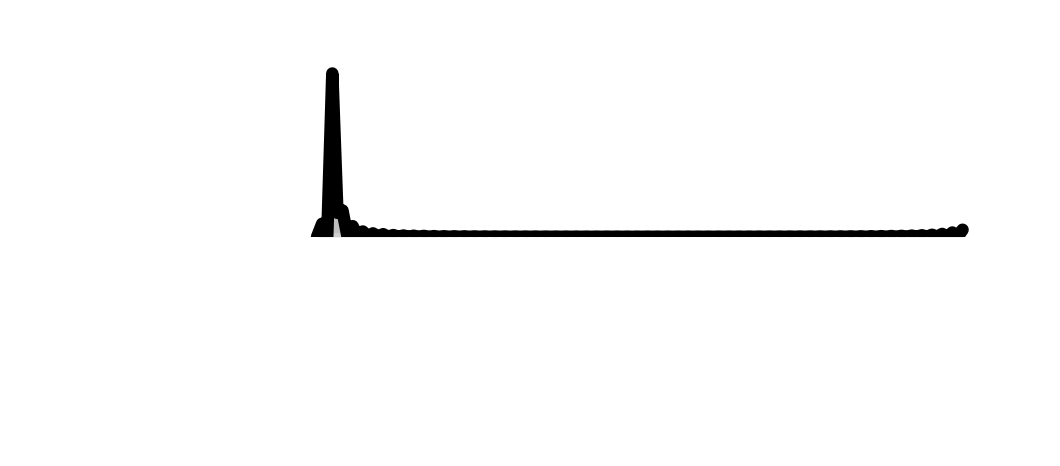}} \\
LIWC Negative affect & 0.0 & 0.13 & 0.03 & 0.03 & \parbox[c]{0.8em}{\includegraphics[width=0.5in]{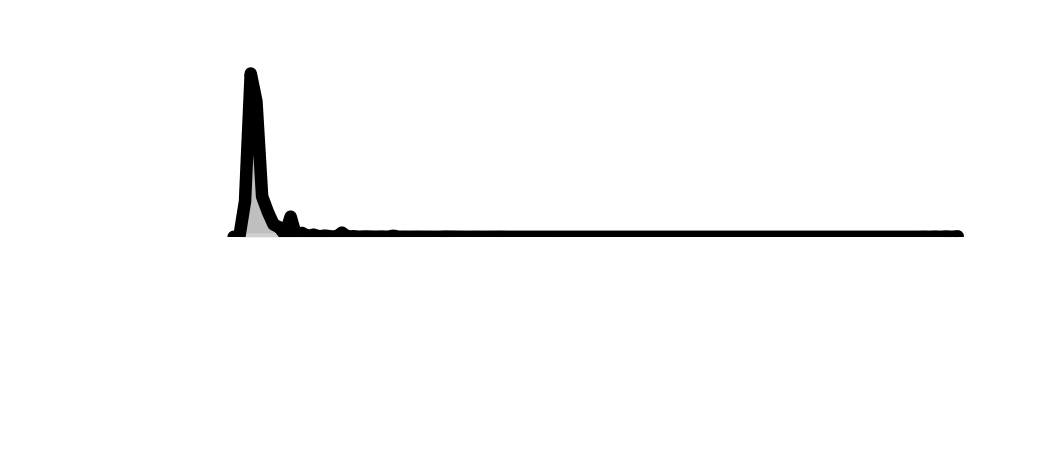}} \\
Coordination in negative affect & 0.0 & 0.26 & 0.06 & 0.06 & \parbox[c]{0.8em}{\includegraphics[width=0.5in]{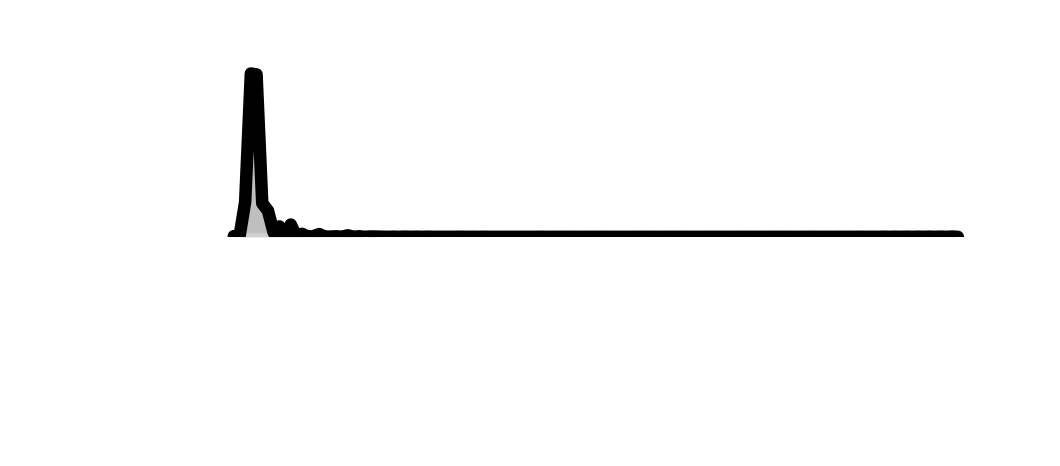}} \\
\textbf{Group Polarization} &  &  &  &  &  \\
First person singular & 0.0  & 0.20 & 0.001 & 0.03 &\parbox[c]{0.8em}{\includegraphics[width=0.5in]{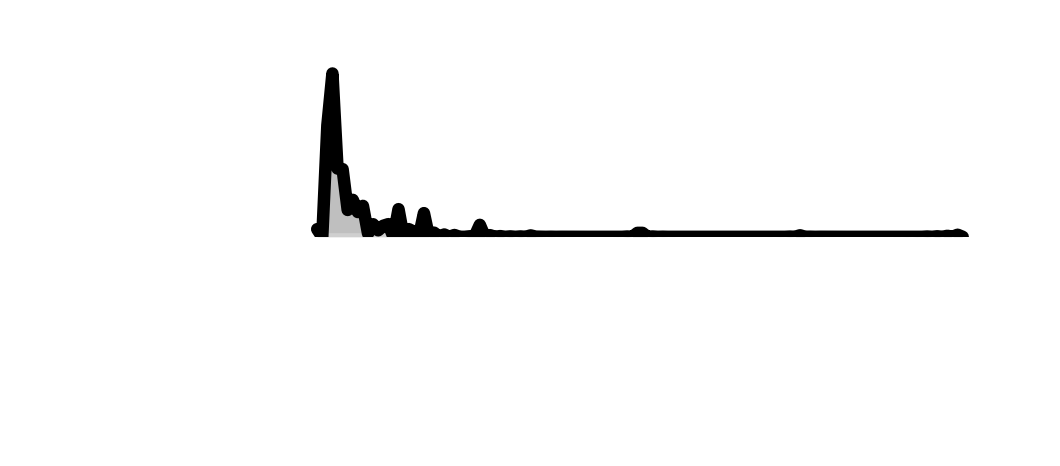}} \\
Coordination in first person singular & 0.0  & 0.16  & 0.002 & 0.01 & \parbox[c]{0.8em}{\includegraphics[width=0.5in]{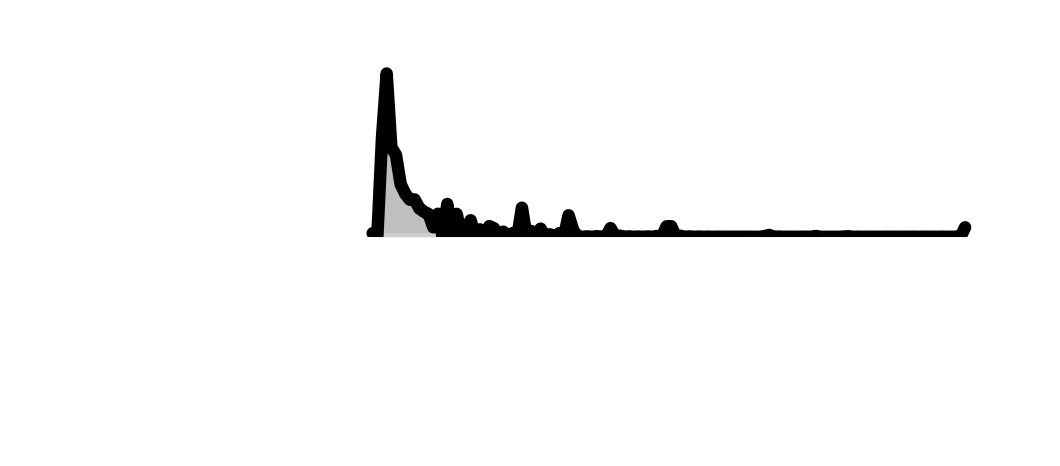}} \\
First person plural & 0.0 & 0.20 & 0.002 & 0.02 &\parbox[c]{0.8em}{\includegraphics[width=0.5in]{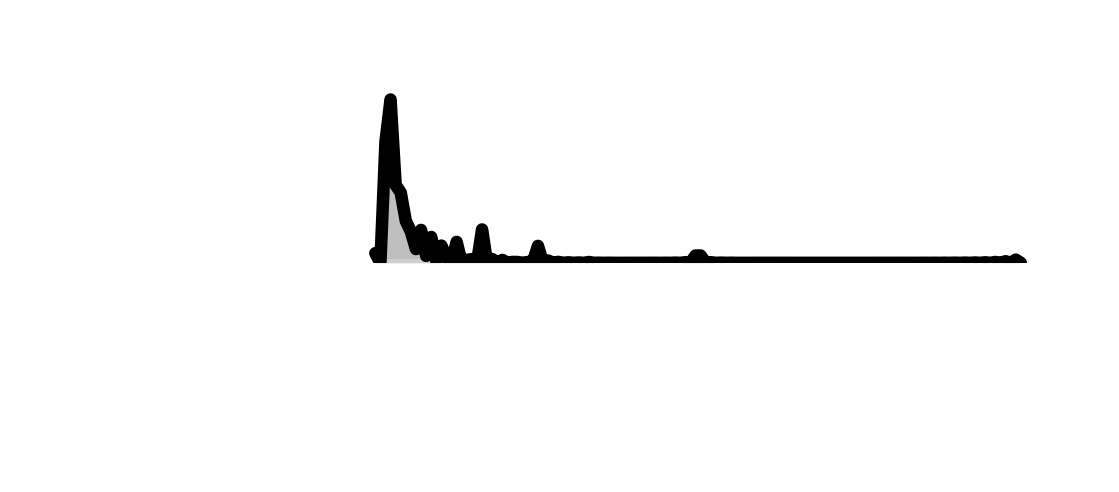}}  \\
Coordination in first person plural & 0.0 & 0.21 & 0.06 & 0.04 & \parbox[c]{0.8em}{\includegraphics[width=0.5in]{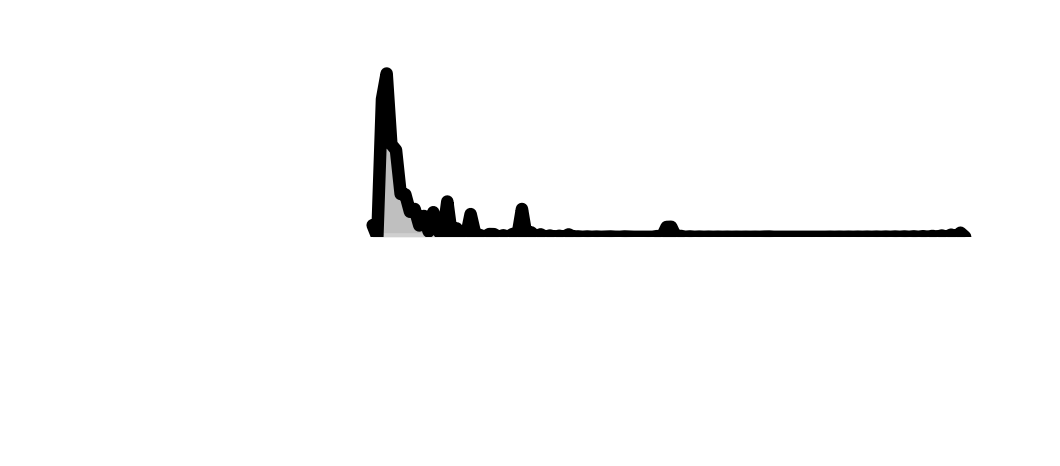}} \\
Second person & 0.0 & 0.07 & 0.004 & 0.04 & \parbox[c]{0.8em}{\includegraphics[width=0.5in]{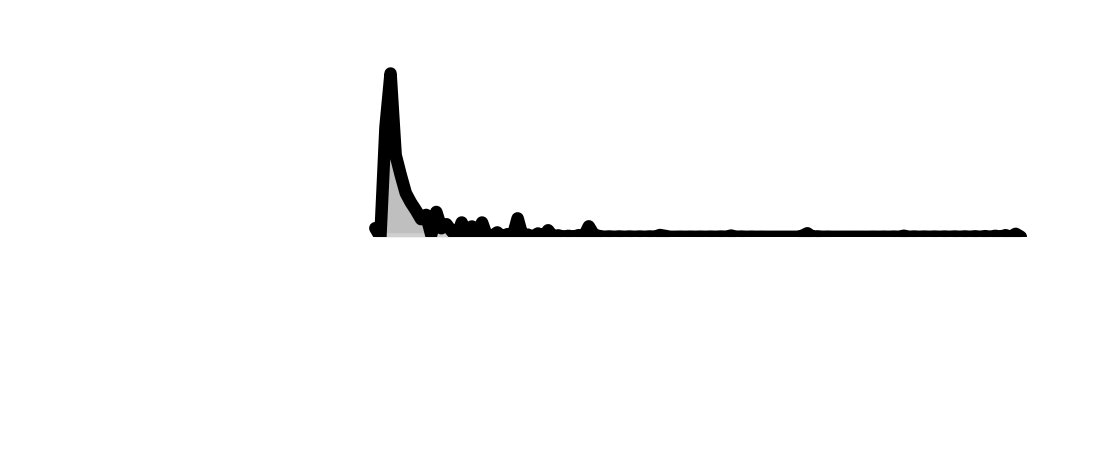}} \\
Coordination in second person & 0.0 & 0.15 & 0.001 & 0.002 & \parbox[c]{0.8em}{\includegraphics[width=0.5in]{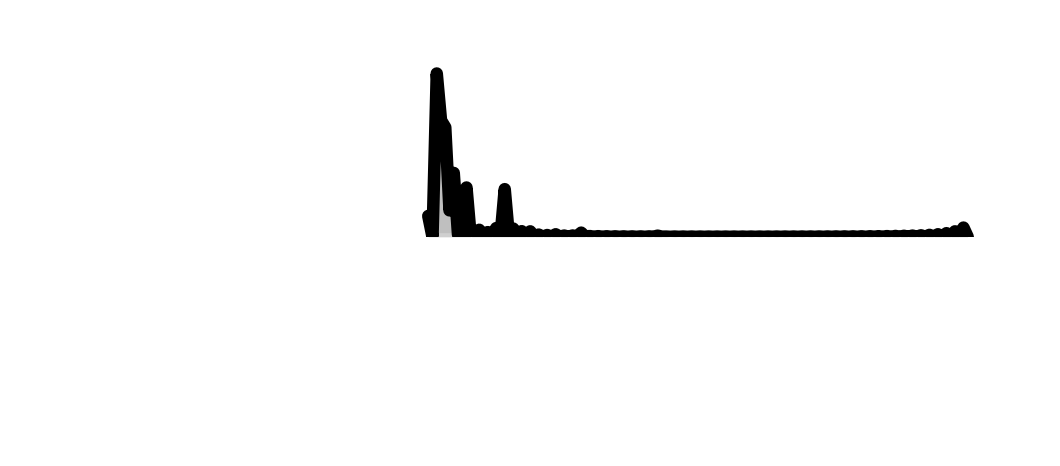}} \\
Third person & 0.0 & 0.30 & 0.01 & 0.02 & \parbox[c]{0.8em}{\includegraphics[width=0.5in]{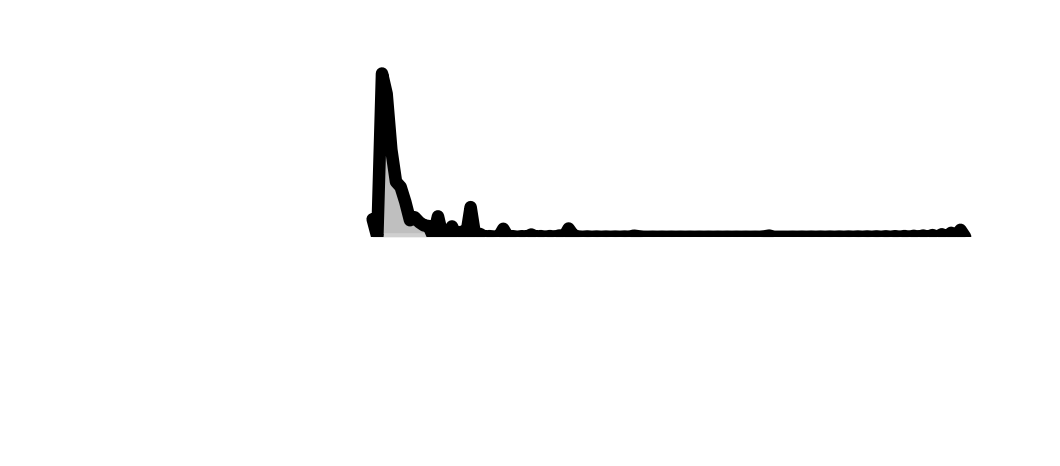}} \\
Coordination in third person & 0.0 & 0.30 & 0.004 & 0.03 &\parbox[c]{0.8em}{\includegraphics[width=0.5in]{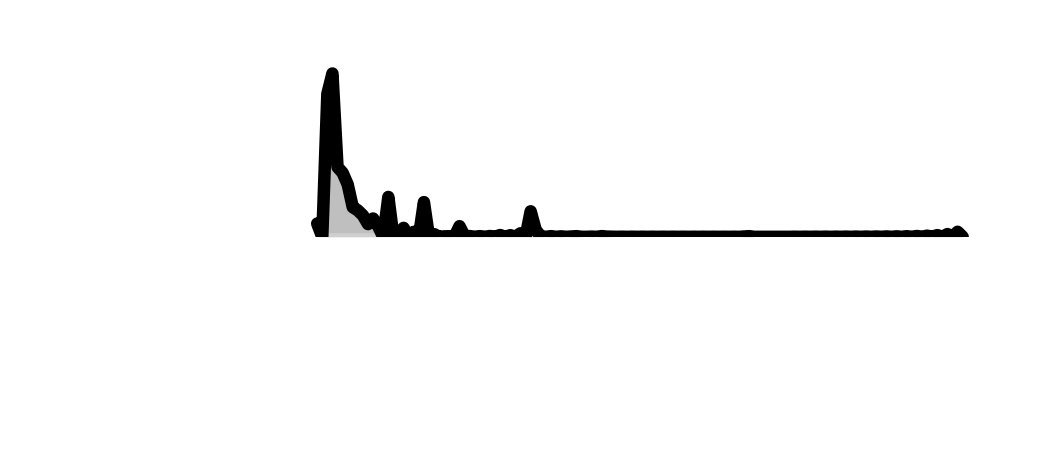}}  \\
\textbf{Self-selection} &  &  &  &  &  \\
Moderated contributions & 0.0 & 1.0 & 0.002 & 0.02 & \parbox[c]{0.8em}{\includegraphics[width=0.5in]{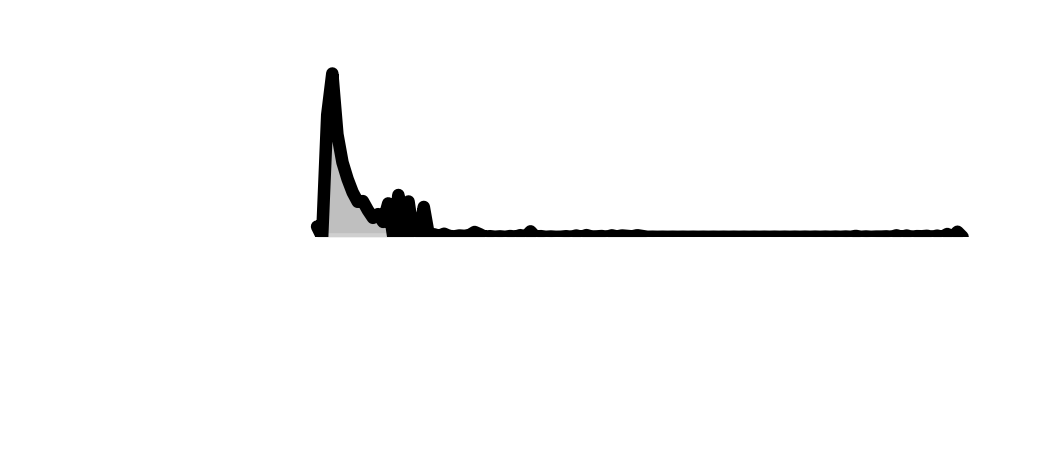}} \\
Negatively scoring contributions & 0.0 & 1.0 & 0.016 & 0.4 & \parbox[c]{0.8em}{\includegraphics[width=0.5in]{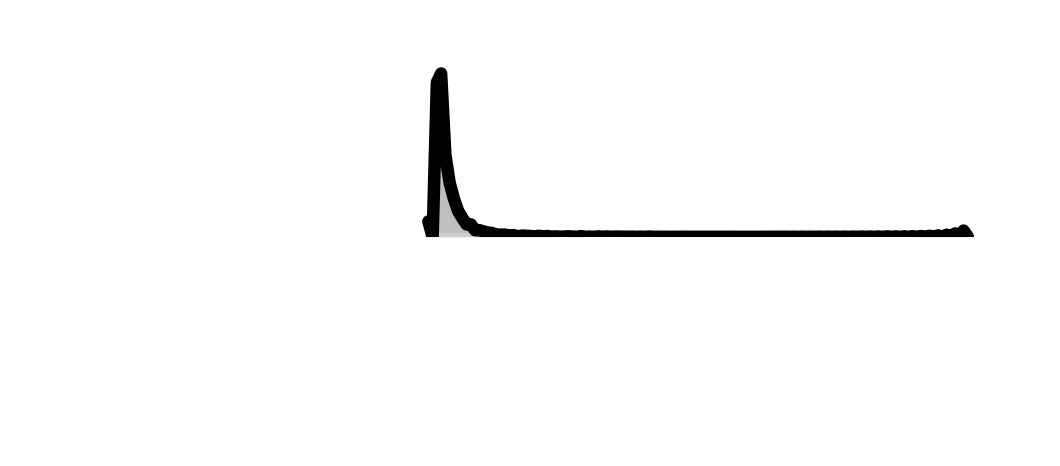}} \\
Contribution trend in the observation period & -1483 & 1371 & 1.78 & 19.99 & \parbox[c]{0.8em}{\includegraphics[width=0.5in]{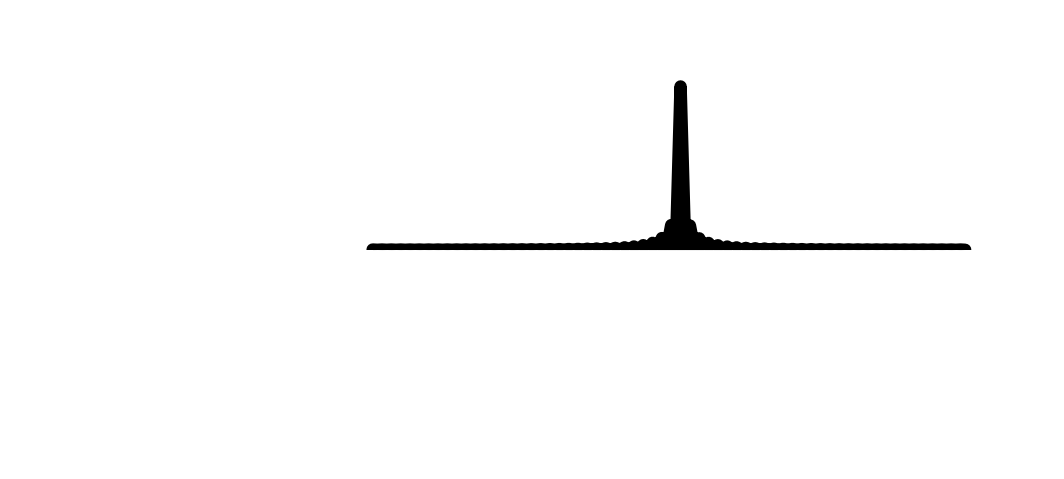}} \\ \bottomrule
\end{tabular}%
\caption{Descriptive statistics and distribution plots for social factors. }
\label{tab:dists_social}
\end{table}

\end{document}